\documentclass[useAMS,usenatbib]{mn2e}

\usepackage{graphics,graphicx}
\usepackage{subfigure}
\usepackage{color}
\usepackage{lscape}
\usepackage{deluxetable}

\begin{document}

\title{Finding Rare AGN: X-ray Number Counts of {\it Chandra} Sources in Stripe 82}

\author[Stephanie M. LaMassa et al.]{Stephanie M. LaMassa$^1$\thanks{E-mail:stephanie.lamassa@yale.edu}, 
C. Megan Urry$^1$, 
Eilat Glikman$^1$, 
Nico Cappelluti$^{2,3}$,  
\newauthor
Francesca Civano$^{4,5}$, 
Andrea Comastri$^2$, 
Ezequiel Treister$^6$, 
Arifin$^7$,
\newauthor
Hans B\"ohringer$^8$, 
Carie Cardamone$^9$, 
Gayoung Chon$^8$,
Miranda Kephart$^1$, 
\newauthor 
Stephen S. Murray$^{10,11}$,
Gordon Richards$^{12}$, 
Nicholas P. Ross$^{13}$,  
Joshua S. Rozner$^1$, 
\newauthor 
Kevin Schawinski$^{14}$\\
$^1$Yale University - Yale Center for Astronomy \& Astrophysics, Physics Department,
$^2$INAF - Osservatorio Astronomico di Bologna,\\ 
$^3$University of Maryland Baltimore College, 
$^4$Harvard-Smithsonian Center for Astrophysics, 
$^5$Dartmouth College, 
$^6$Universidad de Concepci\'on, \\
$^7$National University of Singapore, 
$^8$Max-Planck-Institut F\"ur Extraterrestriche Physik,
$^9$Brown University, 
$^{10}$The Johns Hopkins University,\\
$^{11}$Harvard-Smithsonian Center for Astrophysics,
$^{12}$Drexel University, 
$^{13}$Lawrence Berkeley National Lab,
$^{14}$ETH Z\"urich}  

\maketitle
\begin{abstract}
We present the first results of a wide area X-ray survey within the Sloan Digital Sky Survey (SDSS) Stripe 82, a 300 deg$^2$ region of the sky with a substantial investment in multi-wavelength coverage. We analyzed archival {\it Chandra} observations that cover 6.2 deg$^2$ within Stripe 82 (``Stripe 82 ACX''), reaching 4.5$\sigma$ flux limits of 1.2$\times10^{-15}$, 5.4$\times10^{-15}$ and 2.9$\times10^{-15}$ erg s$^{-1}$ cm$^{-2}$ in the soft (0.5-2 keV), hard (2-7 keV) and full (0.5-7 keV) bands, to find 480, 137 and 705 X-ray sources, respectively. Two hundred fourteen sources are detected only in the full band and 4 sources are detected solely in the soft band. Utilizing data products from the {\it Chandra} Source Catalog, we construct independent Log$N$-Log$S$ relationships, detailing the number density of X-ray sources as a function of flux. The soft and full bands show show general agreement with previous {\it Chandra} surveys; the hard band number counts agree among Stripe 82 ACX, XBootes and XDEEP2, but all 3 are somewhat systematically lower than the counts derived from ChaMP. We compare the luminosity distribution of Stripe 82 ACX with the smaller, deeper CDF-S, E-CDFS and {\it Chandra}-COSMOS surveys, to illustrate the benefit of wide-area surveys in locating high luminosity and/or high redshift AGN. Finally, we compare the identified AGN with predictions from population synthesis models, noting that prior to any spectroscopic follow-up campaign, we have already located roughly half the high luminosity quasars at high redshift expected to lie within the survey area. However, our data also suggests that refinements to population synthesis models will be required.

\end{abstract}

\section{Introduction}

Supermassive black holes (SMBHs) found in the centers of all massive galaxies grow by accreting matter, and are dubbed Active Galactic Nuclei (AGN) during this phase. X-rays are the most effective way to find moderate to high-luminosity AGN because they emit a significant fraction of their bolometric luminosity at these energies compared with inactive galaxies and nearly all X-ray point sources with L$_{x} > 10^{42}$ erg/s are AGN \citep[e.g.][]{Persic}. (This dividing line misses low luminosity accreting black holes and can potentially include a few very energetic starburst galaxies that might attain these X-ray luminosities.) Optical studies of the quasar luminosity function (QLFs) from large areas covered by the Sloan Digital Sky Survey (SDSS) have indicated that black hole growth evolves as a function of luminosity \citep[e.g.,][]{Richards, Ross}, although these unobscured AGN likely represent only $\sim$30\% of the total black hole growth \citep[e.g.,][]{treister}.

The deepest X-ray surveys to date have been successful in elucidating the low to moderate luminosity AGN population. The {\it Chandra} Deep Fields North and South (CDF-N and CDF-S) have peered to depths of 2 Ms and 4 Ms, reaching to 0.5-2 keV (2.0-8.0 keV) flux limits of 2.5$\times10^{-17}$ (1.4$\times10^{-16}$) erg s$^{-1}$ cm$^{-2}$ and 9.1$\times10^{-18}$ (5.5$\times10^{-17}$) erg s$^{-1}$ cm$^{-2}$ respectively, in CDF-N and CDF-S \citep{cdfn,cdfs}. The Extended {\it Chandra} Deep Field South (ECDF-S) surrounded the original CDF-S survey, covering $\sim$0.3 deg$^2$ with sensitivity limits of $\sim 1.1\times 10^{-16}$ and $\sim 6.7\times10^{-16}$ erg cm$^{-2}$ s$^{-1}$ in the 0.5-2 keV and 2.0-8.0 keV bands respectively \citep{leh-ecdfs}. These observations are integral components of the larger multi-wavelength surveys of Great Observatories Origins Deep Survey \citep[GOODS;][]{giv-goods,goods} and Multi-wavelength Survey by Yale-Chile \citep[MUSYC;][]{musyc}. {\it XMM-Newton} and {\it Chandra} have done deep surveys of the COSMOS field, reaching 0.5-2 keV fluxes down to 7$\times10^{-16}$ erg cm$^{-2}$ s$^{-1}$ on the whole field \citep{xmmcosmos,xmm-cap} and 1.9$\times10^{-16}$ erg cm$^{-2}$ s$^{-1}$ in the inner area \citep{C-Cosmos,civano2}, respectively. In the next two years, {\it Chandra} will survey the whole COSMOS field to the depth now available only in the central area (COSMOS Legacy, PI: F. Civano). However, since these surveys cover relatively small areas, high luminosity and/or high redshift AGN remain under-sampled compared with the population that would be detected in much wider areas. As half of the black hole growth occurs in high luminosity AGN \citep{treister12}, the existing census of black hole growth over cosmic time is incomplete. Indeed, X-ray QLFs at high redshift ($z > 3$) are only constrained at lower luminosities, necessitating optical studies to fill in the high luminosity parameter space to allow the QLF to be fit over several decades of luminosity \citep[e.g.,][]{fiore}. Such a method assumes optically selected quasars represent the same population as X-ray selected AGN, which may be incorrect. Indeed, this assumption is incorrect at lower luminosities \citep[e.g.,][]{goods}.

Larger volume X-ray surveys are necessary to sample a representative population of growing black holes since high redshift and high luminosity sources are relatively rare. So far, only  XBo\"otes \citep[9 deg$^2$,][]{murray, kenter}, ChaMP,  a serendipitous survey based on archival {\it Chandra} data \citep[10 deg$^2$,][]{champ1, champ2}, extended ChaMP/SDSS \citep[$\sim$32 deg$^2$,][]{green} and the {\it XMM-Newton} Large Scale Structure \citep[XMM-LSS, $\sim$11 deg$^{2}$, with a full survey coverage of 50 deg$^2$ planned,][]{lss1,lss2} surveys come close to exploring this part of parameter space. Ancillary multi-wavelength data are then needed to provide additional critical information about these sources, such as optical identifications, with redshifts, to accurately calculate AGN and host galaxy properties like accretion luminosity, galaxy morphology, stellar content, star formation activity, galaxy mass, and so on. Follow-up spectroscopy over such large areas have been obtained only very slowly. About half of the X-ray sources in XBo\"otes have MMT Hectospec spectra as part of the AGN and Galaxy Evolution Survey \citep[AGES,][]{kochanek}, but this campaign took over 4 years. The optical follow-up for ChaMP also required a similar time investment \citep{trichas}. Though \citet{mateos} produced the Log$N$-Log$S$ distribution for the full {\it XMM-Newton} Serendipitous Source Catalog, covering $\sim$130 deg$^2$, multi-wavelength follow-up observations over this area to identify these X-ray sources and to produce an accurate quasar luminosity function is prohibitively time-consuming.

To efficiently probe the high luminosity and high redshift AGN population, we have begun a large area X-ray survey in SDSS Stripe 82, a $\sim$300 deg$^2$ area in a region of space that harbors substantial observational investments from an assortment of multiwavelength facilities, including FIRST and EVLA \citep[radio,][]{Becker, Hodge}; Atacama Cosmology Telescope \citep[millimeter,][]{das}; UKIDSS and VISTA Hemisphere Survey \citep[near-infrared,][]{ukidss, vista} and GALEX \citep[ultraviolet,][]{galex}. Stripe 82 extends from -60$^{\circ}<$RA$< $60$^{\circ}$ and -1.25$<$Dec$<$1.25, and is thus accessible from both hemispheres. An advantage of Stripe 82 over the full SDSS catalog, and by extension the ChaMP/SDSS survey, is the increased depth in the optical: the imaging in Stripe 82 reaches two magnitudes deeper than any individual SDSS scan since each individual area was imaged $\sim$80 times. The optical spectral density of Stripe 82 is also exceptionally high thanks to observations with SDSS, 2dF and WiggleZ \citep{wigglez}, and continues to increase with spectroscopic campaigns from SDSS-III BOSS \citep{sdssiii} and AUS (Croom et al., in prep.): at present, $>$400 optical objects per square degree have spectra.  

Combining archival {\it XMM-Newton} and {\it Chandra} observations as well as newly obtained {\it XMM-Newton} observations (which we will report in a future publication due to the complicated data reduction procedures required for these {\it XMM-Newton} data), the total X-ray coverage of Stripe 82 (Stripe 82 X) reaches $\sim$17 deg$^2$ so far. Taking advantage of the high level of optical spectroscopic completeness, this survey will reveal more rare high luminosity (L $> 3\times 10^{44}$ erg s$^{-1}$) AGN at all redshifts and in the distant universe ($z > 2$) than previous X-ray surveys. 

In this pilot study, we present X-ray source counts from a new large-area Archival Chandra survey in Stripe 82 (hereafter, ``Stripe 82 ACX''), covering a largely non-contiguous area of $\sim$6.2 deg$^2$, and make comparisons for the first time to the other three largest Chandra surveys. We also describe the general characteristics of AGN found in Stripe 82 ACX. Though the full 17 deg$^2$ data will be required to generate the high end of the X-ray quasar luminosity function, the present number counts - and in particular, the area-flux curve - are essential for deriving the QLF. We demonstrate that these number counts do indeed sample the high luminosity AGN population under-represented in smaller area surveys. Furthermore, our results suggest that current population synthesis models are incorrect at high luminosity and/or high redshift. Throughout the paper, we use cosmology H$_{0}$ = 70 km/s/Mpc, $\Omega_M$ = 0.27 and $\Lambda$=0.73.

\section{Stripe 82 ACX: Archival {\it Chandra} Data}

We utilized the {\it Chandra} Source Catalog \citep{CSC} to identify X-ray sources and estimate the survey sensitivity of Stripe 82 ACX. The {\it Chandra} Source Catalog currently contains point and compact ($\leq30^{\prime\prime}$) X-ray sources from the first eight years of public {\it Chandra} imaging observations, with associated data products, such as events files, exposure maps, etc., reduced homogeneously. Full details of the data reprocessing are provided in \citet{CSC}. In brief, all observations for the {\it Chandra} Source Catalog were recalibrated with a calibration database (CALDB) version created specifically for the {\it Chandra} Source Catalog. ACIS observations had the time-dependent gain calibration applied and were corrected for CTI (CCD charge transfer inefficiency). The CIAO task {\em acis\_run\_hotpix} was run to remove pixel afterglow events, and bad pixels and hot pixels were flagged for removal. The background was screened to reduce the non X-ray background, removing intervals with strong background flaring. 

For a source to be included in the {\it Chandra} Source Catalog, a conservative threshold was chosen to mitigate the inclusion of spurious detections as real sources. Source detection was performed by the CIAO task {\em wavdetect}, with a limiting significance level of 2.5$\times10^{-7}$, corresponding to $\sim$1 false detection per pointing.  An additional cut on flux significance was then applied. The {\it Chandra} Source Catalog estimates this significance by $S/\sigma_e$, where $\sigma_e = FWHM/(2\sqrt{2ln2})$ and the FWHM is of the posterior probability density ($p(S|CB)$) for the flux ($S$) assuming that the total counts ($C$) and background counts ($B$) have a Poisson distribution.  The flux had to be significant at $\geq$3-$\sigma$ level in at least one flux band, generally corresponding to $\sim$10 source photons on-axis, increasing to $\sim$20-30 source counts off-axis since the {\it Chandra} PSF degrades with off-axis distance. This threshold is more conservative than those used in previous {\it Chandra} surveys \citep[see][for a comparison]{CSC, deep2} and removes $\sim$1/3 of the sources detected by {\em wavdetect}. Though this high signal-to-noise cut makes Stripe 82 ACX incomplete at low fluxes, this does not adversely affect the science goals of this project, which are to probe the high X-ray luminosity AGN population and to accurately constrain their number density and evolution. Surveys such as CDF-S \citep{cdfs} are optimized to uncover X-ray sources at the faintest levels, whereas the wide survey area of Stripe 82 X complements this parameter space by better exploring the high X-ray flux regime.

\begin{figure}
\centering
{\includegraphics[scale=0.75,angle=90]{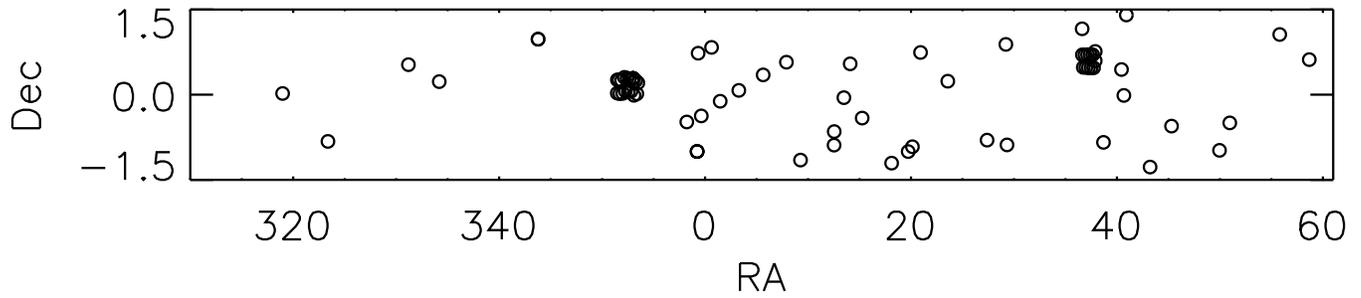}}
\caption[]{\label{pointings} Locations of archival {\it Chandra} observations in Stripe 82. The dense cluster of pointings at RA = $\sim$352$^{\circ}$ and RA = $\sim$37$^{\circ}$ are Fields 3 and 4, respectively, of the XDEEP2 Survey \citep[see][for details, including catalog of X-ray sources and optical counterparts]{deep2}.}

\end{figure}

Figure \ref{pointings} shows the positions of the archival {\it Chandra} observations in Stripe 82. We note that fields targeting galaxy clusters in Stripe 82 were {\it a priori} removed from our analysis. We note that including galaxy cluster fields would make no noticeable differences on our derived Log$N$-Log$S$ relationships. The two dense pointing regions at RA = $\sim$352$^{\circ}$ and RA = $\sim$37$^{\circ}$ are observations from the XDEEP2 survey, Fields 3 and 4, respectively \citep{deep2}.

\renewcommand{\thefootnote}{\fnsymbol{footnote}}

\subsection{\label{sens}Survey Sensitivity}
We estimated the Stripe 82 ACX survey sensitivity at the 4.5$\sigma$ flux significance threshold, corresponding to $\sim$1 spurious source per 100 deg$^{2}$\footnote{We estimated this number by noting that $\sim$1100 unique sources were detected at all significance levels by the {\it Chandra} Source Catalog in the 6.2 deg$^2$ covered here and then extrapolating to a 100 deg$^2$ area.}, by creating sensitivity maps for each observation in the soft (0.5-2 keV), hard (2-7 keV) and full (0.5-7 keV) bands. These sensitivity maps give the limiting flux at the cited significance threshold for each pixel in the detector. We note that the {\it Chandra} Source Catalog runs the source detection algorithm and provides products and fluxes in the more narrow ``s'' (0.5-1.2 keV) and ``m'' (1.2-2 keV) bands. For consistency with previous X-ray surveys, however, we explore the {\it Chandra} number counts in the traditional ``soft'' band, 0.5-2 keV. The hard and full bands defined here are consistent with those from the {\it Chandra} Source Catalog, i.e., only the ``soft'' band is redefined.

First, exposure maps were created, where we used the level 3 event files from the {\it Chandra} Source Catalog and downloaded the aspect solution file, mask file and parameter block file from the {\it Chandra} archive; these level 3 event files are similar to the level 2 event files included in the {\it Chandra} data archive (i.e., filtered files suitable for data analysis), but with different good time interval (GTI) filters applied to remedy background flaring. We created aspect histograms for each active chip in an observation with CIAO tool {\em asphist}, using the aspect solution and event files as input. We then created an instrument map for each chip, which provides information about the effective area as a function of detector position, using the CIAO tool {\em mkinstmap}. As the effective area is a function of energy, we provided an input file of spectral weights derived from the CIAO routine {\em make\_instmap\_weights} using a power law model with $\Gamma$ = 1.4, to be consistent with previous {\it Chandra} surveys to which we compare our number counts (i.e., ECDF-S, {\it Chandra}-COSMOS, XDEEP2 and ChaMP). {\em mkinstmap} was also given the mask file and parameter block file, which defines the clocking parameters for each pixel (i.e., how long the pixel is exposed before read-out). Using these instrument maps and aspect histograms, an exposure map for each chip is then created using CIAO tool {\em mkexpmap}. Finally, these individual chip by chip exposure maps are combined into one image  with CIAO routine {\em reproject\_image}, producing an exposure map for each observation in each band.

To create the sensitivity maps, we used the background images in the hard and full bands provided by the {\it Chandra} Source Catalog, with the narrow ``s'' and ``m'' bands added to produce a soft band background image. Using CIAO tool {\em mkpsfmap}, new PSF images were then generated at the spectral weighted mean energies of each band, i.e., 1.02 keV, 3.79 keV and 1.98 keV for the soft, hard and full bands, respectively, again assuming a powerlaw with $\Gamma$=1.4. Finally, sensitivity maps for each obsid were created from these exposure maps and PSF images, with CIAO tool {\em lim\_sens}, using a 4.5$\sigma$ flux threshold. These sensitivity maps were then converted from photon flux units to energy units using an absorbed power law model with $\Gamma$ = 1.4 and N$_{H}$ = 3$\times 10^{20}$ cm$^{-2}$, the approximate absorption through our Galaxy in the direction of Stripe 82 (see Table \ref{info} for conversion factors).

\begin{figure}
\centering
{\includegraphics[scale=0.5,angle=90]{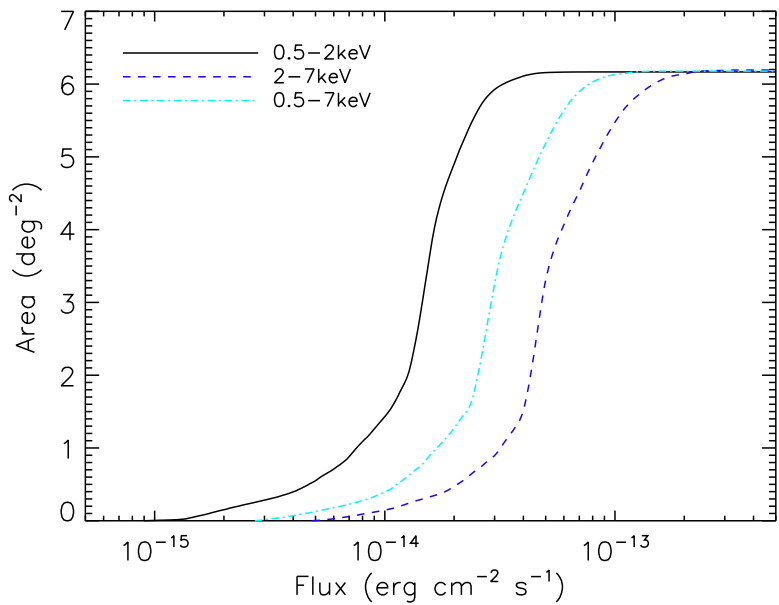}}
\caption[]{\label{area-flux} Stripe 82 ACX survey area as a function of limiting flux for the soft (0.5-2 keV, solid dark line), hard (2-7 keV, dashed blue line) and full (0.5-7 keV, dotted dashed cyan line) energy bands. Details concerning the derivation of these curves are given in the text.}

\end{figure}

We effectively mask out the chip edges where source detection becomes improbable by ``turning off'' pixels where the exposure map drops below 15\% of the maximum value. We note that no sources fall within these masked out pixels. To gauge full sensitivity across the survey, the individual sensitivity maps were overlaid on a grid of pixels spanning the Stripe 82 area. In overlapping regions, the most sensitive pixel (i.e., the lowest limiting flux) was chosen. From this Stripe 82 region sensitivity map, we then calculated the cumulative histogram of survey area as a function of limiting flux, producing the area-flux curves in Figure \ref{area-flux}. Stripe 82 ACX covers $\sim$6.2 deg$^2$ of non-overlapping area, with cluster fields removed, which would add an additional 1.3 deg$^2$ of coverage. As noted in Table \ref{info}, the approximate flux limits of this survey are 1.2$\times10^{-15}$, 5.4$\times10^{-15}$ and  2.9$\times10^{-15}$ erg cm$^{-2}$ s$^{-1}$, with half the survey area visible to approximate depths of 1.5$\times10^{-14}$, 4.8$\times10^{-14}$ and 2.9$\times10^{-14}$ erg cm$^{-2}$ s$^{-1}$ in the soft, hard and full bands, respectively.

\subsection{\label{sel}Source Selection}
\renewcommand{\thefootnote}{\arabic{footnote}}

From the Source Observation Table in the {\it Chandra} Source Catalog, we isolated point sources lying in Stripe 82 that are not saturated, on the ACIS read-out streak, or suffering from pile-up.\footnote{That is, we chose sources where the following flags were set to 0: $extent\_code$, $sat\_src\_flag$, $streak\_src\_flag$, and where the pileup fraction was less than 10\%. A source is considered extended (i.e., $extent\_code$ is set to 1) if the intrinsic source extent, found by deconvolving the local PSF from the observed source extent (parameterized as a rotated elliptical Gaussian), is inconsistent with a point source at the 90\% confidence level. Only one source suffered from significant pileup (i.e., pileup fraction 25\%) and was thus discarded from our source list.} For the hard and full bands, the aperture corrected flux in photon units was converted to energy units using the conversion factors listed in Table \ref{info}. To derive the soft flux, we had to ensure the ``s'' and ``m'' fluxes were combined with appropriate weights to match that of the sensitivity map. The aperture corrected net counts in the ``s'' ($C_s$) and ``m'' ($C_m$) bands from the {\it Chandra} Source Catalog were added, giving soft band counts ($C_{soft}$):
\begin{equation} C_{soft} = C_s + C_m. \end{equation}
These counts were then divided by the pixel from the soft band exposure map corresponding to the location where the source was detected ($exp_{soft}$), where the exposure map was created as described in Section \ref{sens}:
\begin{equation} F^{ph}_{soft} = C_{soft} / exp_{soft}.  \end{equation}
Again, the soft flux in photon units were converted to energy units via the conversion factor listed in Table \ref{info}. In practice, a source could be detected with {\it wavdetect} run in the full soft band, but not in the more narrow ``s'' and ``m'' bands. However, the {\it Chandra} Source Catalog includes a flux measurement for each band as long as the source was detected at the 3$\sigma$ level in any given band and we therefore have enough information to calculate the soft band flux and then test its significance.  As we further impose a 4.5$\sigma$ threshold for inclusion of a source in the Stripe 82 ACX catalog, we expect that missed objects, sources that would have been detected at the 4.5$\sigma$ level in the 0.5-2 keV band but not at the 3$\sigma$ level in any individual band, are rare.

To determine whether a detection is significant at the 4.5$\sigma$ level, and for consistency with the area-flux curves from which we generate the Log$N$ - Log$S$ relationships, we compared the X-ray source list with the sensitivity maps: the source flux had to meet or exceed the limiting flux from the sensitivity map at the pixel where the source was identified in order for us to consider the source as significantly detected in that specific energy band. There were 112 sources detected in multiple observations (identified by the {\em msid} flag in the Master Source Observation Table). For these objects we chose the flux corresponding to the most sensitive observation. 

To avoid skewing our statistics and introducing a bias into our catalog, we also removed targeted sources from our catalog, identified as the source within 5$^{\prime\prime}$ of the user supplied RA and Dec for the target of the observation; we note that only 19 pointings from the 73 observations in this analysis had sources that met this criterion and were subsequently removed. As mentioned previously, 29 pointings are from the XDEEP2 survey, and thus do not have targeted sources. For the remaining 25 observations, either the targeted object was not detected at the 4.5$\sigma$ level or the observer supplied X-ray coordinates of the target were not well determined. 

Though we {\em a priori} removed observations that targeted galaxy clusters and groups, the fields that targeted specific objects could potentially be biased by clustering. To test if such an effect factors into our sample, we compared the spectroscopic redshifts for sources detected serendipitously in each field with the redshift of the targeted source (see Section \ref{opt_info} for details on matching X-ray sources with optical counterparts and finding associated spectroscopic redshifts). In 14 of the 19 fields where the targeted source was detected at a significant level and thus removed, the spectroscopic redshifts of the target and more than one X-ray field source were available. We find that in all cases, these serendipitous sources are not associated with the target: the dispersion between the target redshift and those in the field range from 0.32 to 3.13. Bias due to source clustering around targeted objects therefore does not affect this survey.

In total, we detect 709 unique X-ray sources in Stripe 82 ACX, with 705 detected in the full band, 480 in the soft band and 137 in the hard band at the 4.5$\sigma$ level. Of these, 4 objects were detected in only the soft band and 214 were detected solely in the full band; none were detected in just the hard band.

\section{Stripe 82 ACX Number Counts}
We present the number density of sources as a function of flux, i.e., the log$N$ - log$S$ relation. The binned differential number counts are given by:

\begin{equation} \frac{dN}{dS} = \frac{\Sigma^{i=n}_{i=1} \frac{1}{\Omega_i}}{\Delta S_j};\end{equation}
where n is the number of sources in bin $j$, $\Omega_i$ is the limiting sky coverage associated with the $i$th source and $\Delta S_j$ is the width of the flux bin. Here, we follow the prescription of \citet{mateos} to calculate $S_j$, a weighted flux to represent the bin centroid:
\begin{equation} S_j = \Sigma^{i=n}_{i=1} w_i \times S_i, w_i = \frac{\frac{1}{\Omega_i}}{\Sigma_{i=1}^{i=n}\frac{1}{\Omega_i}}.\end{equation}
The error is given by Poissonian statistics:
\begin{equation} \frac{dN/dS}{\sqrt{n}}. \end{equation}
We have binned by 20, 10 and 30 sources in the soft, hard and full bands, respectively, so that each bin, except for the highest flux bin, has an equal number of sources rather than equal flux widths.

In integral form, the cumulative source distribution is represented by:

\begin{equation} N(>S) = \sum_{i=1}^{N_s}  \frac{1}{\Omega_i}, \end{equation}
where N($>$S) is the number of sources with a flux greater than $S$ and $\Omega_{i}$ is defined as above. The associated error is the variance:

\begin{equation} \sigma^2 = \sum_{i=1}^{N_s}  (\frac{1}{\Omega_i})^2. \end{equation}

\begin{figure}
\centering
\subfigure[]{\includegraphics[scale=0.35,angle=90]{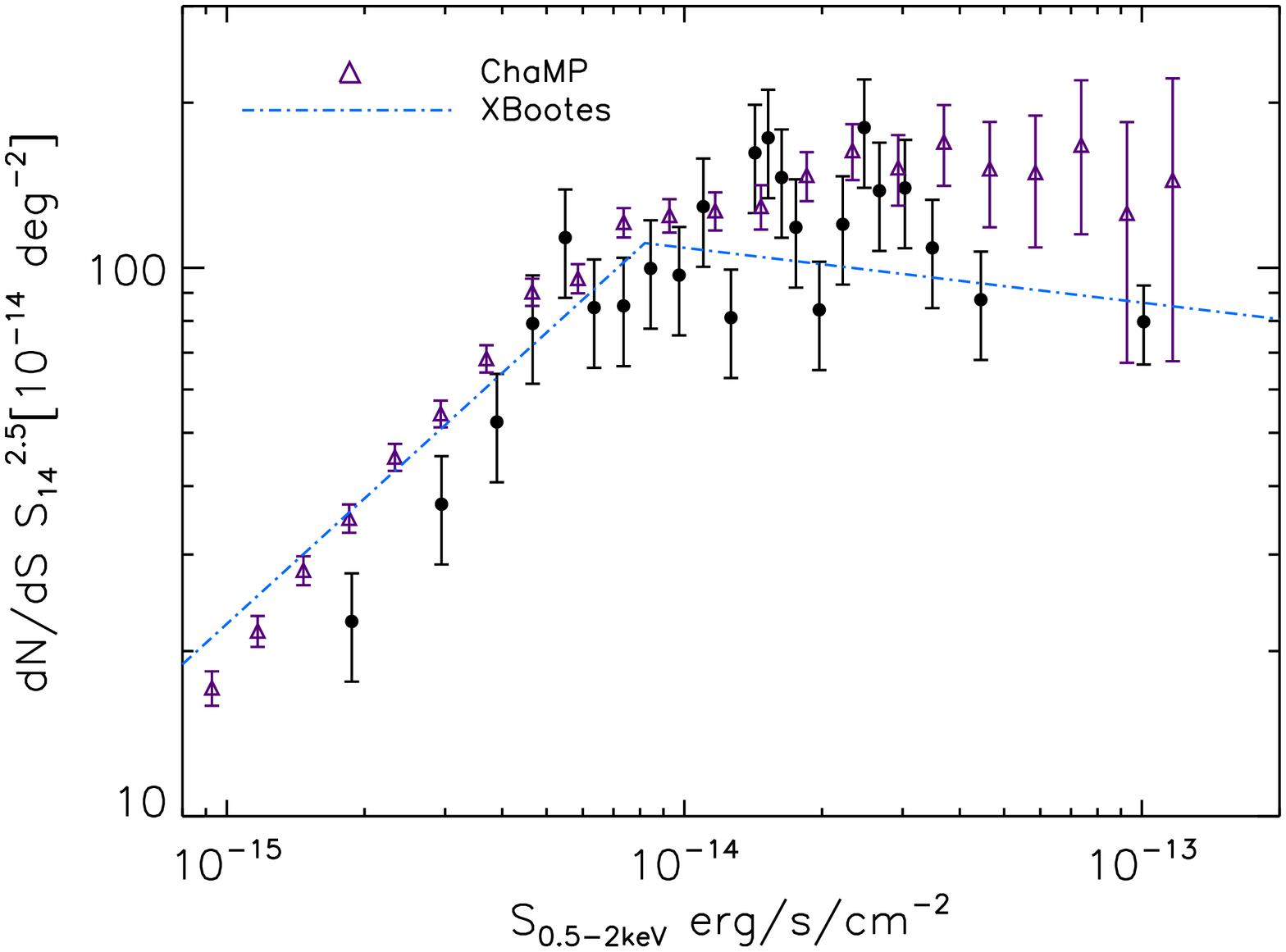}}~
\subfigure[]{\includegraphics[scale=0.35,angle=90]{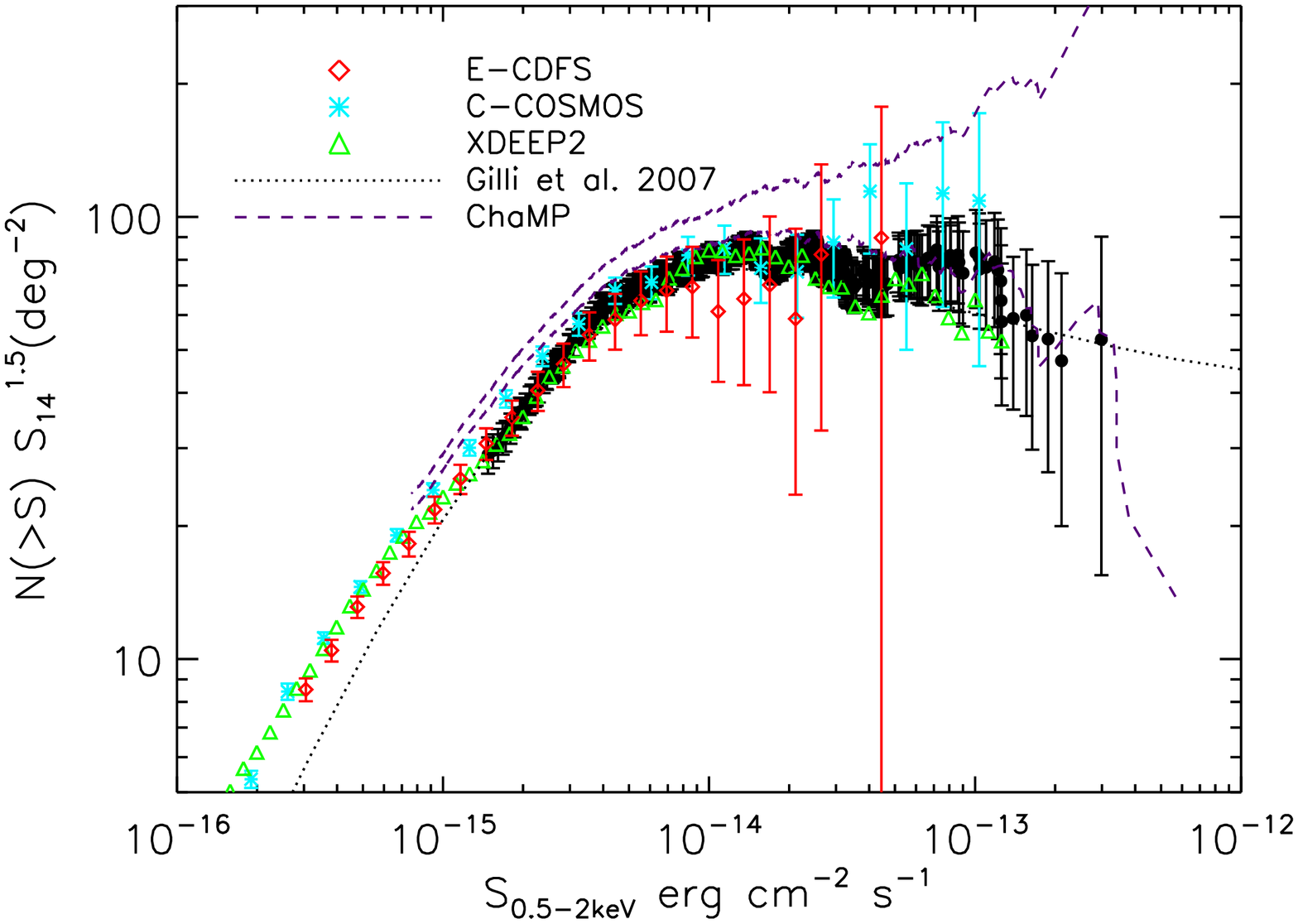}}
\subfigure[]{\includegraphics[scale=0.35,angle=90]{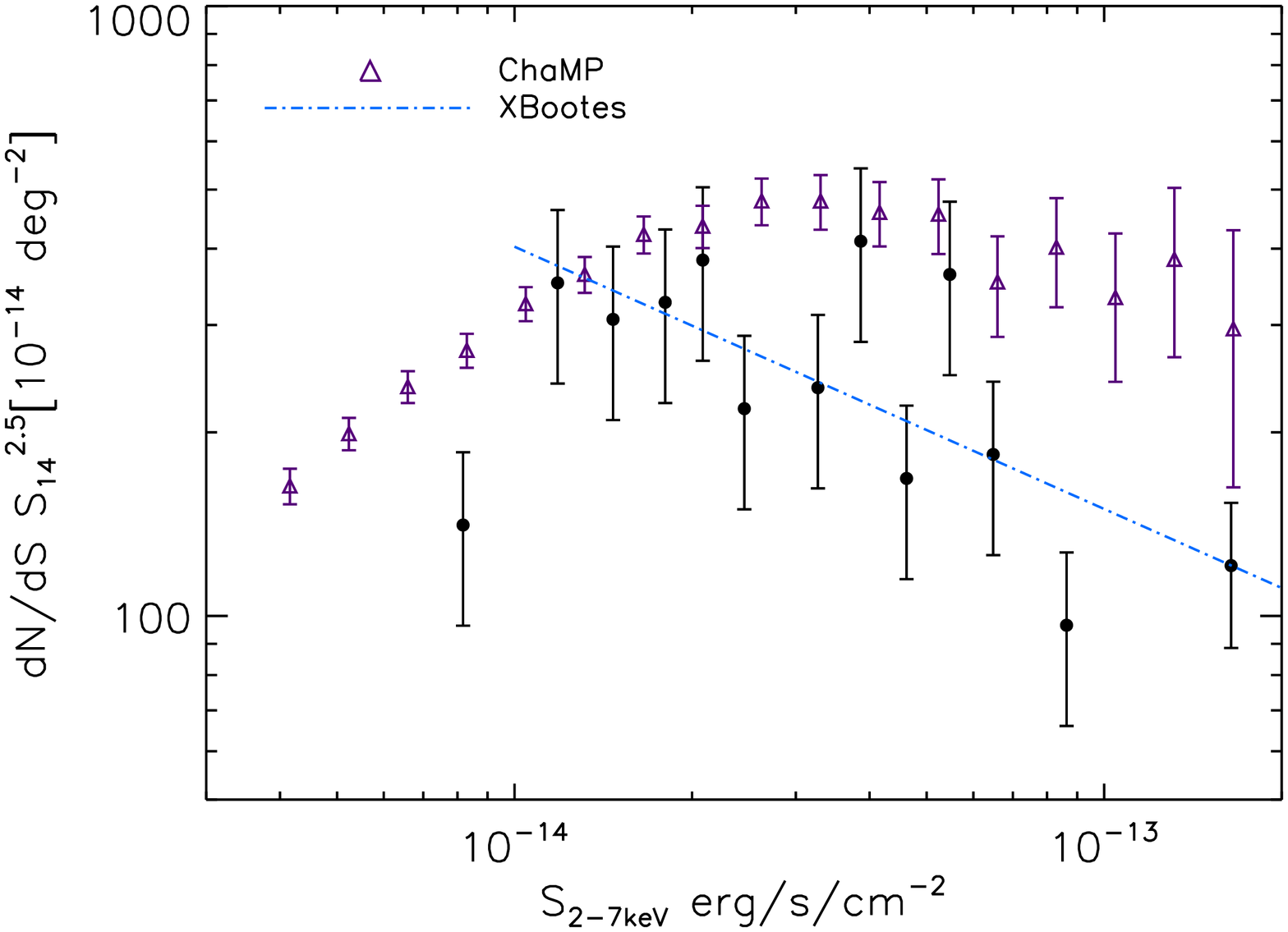}}~
\subfigure[]{\includegraphics[scale=0.35,angle=90]{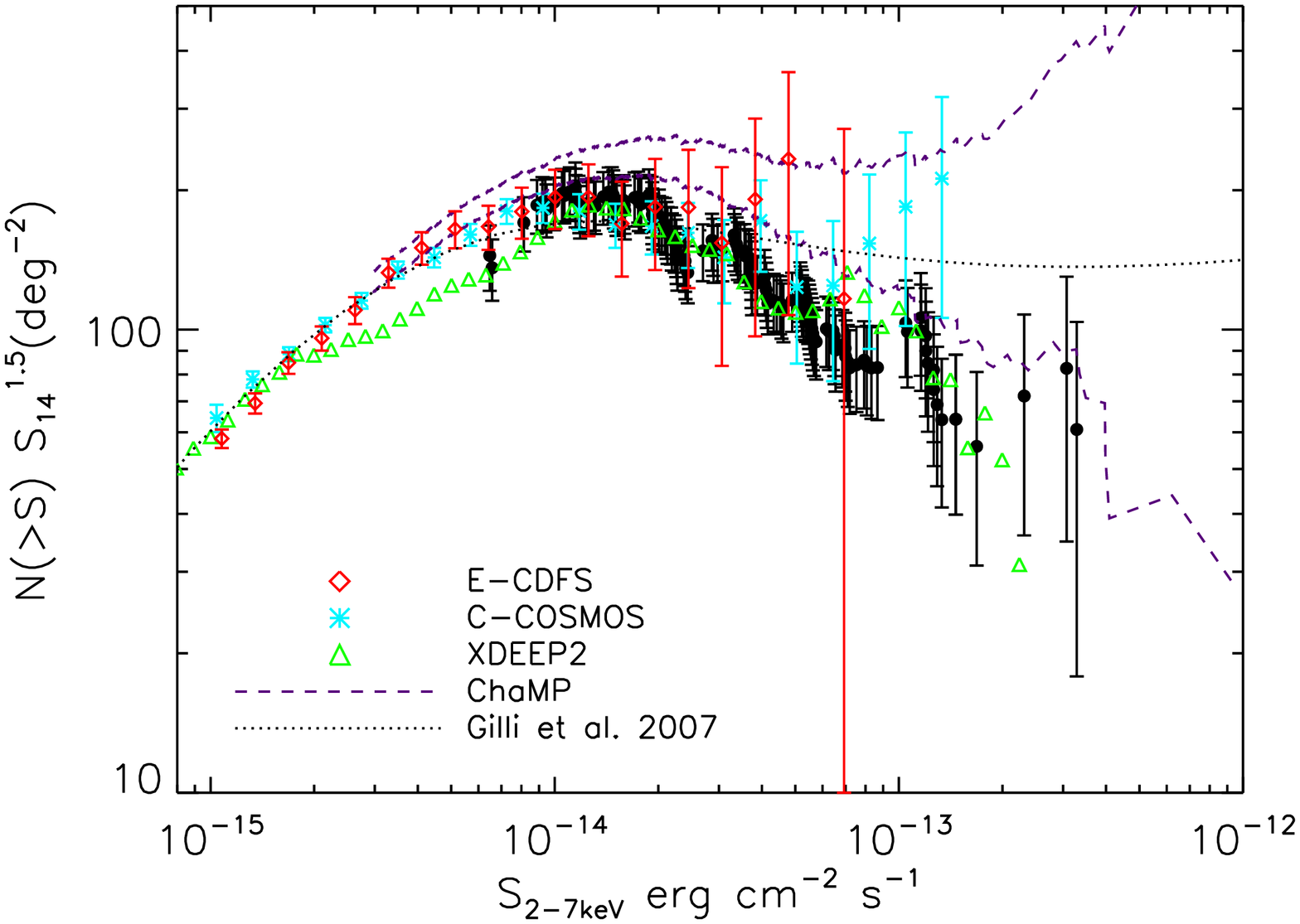}}
\subfigure[]{\includegraphics[scale=0.35,angle=90]{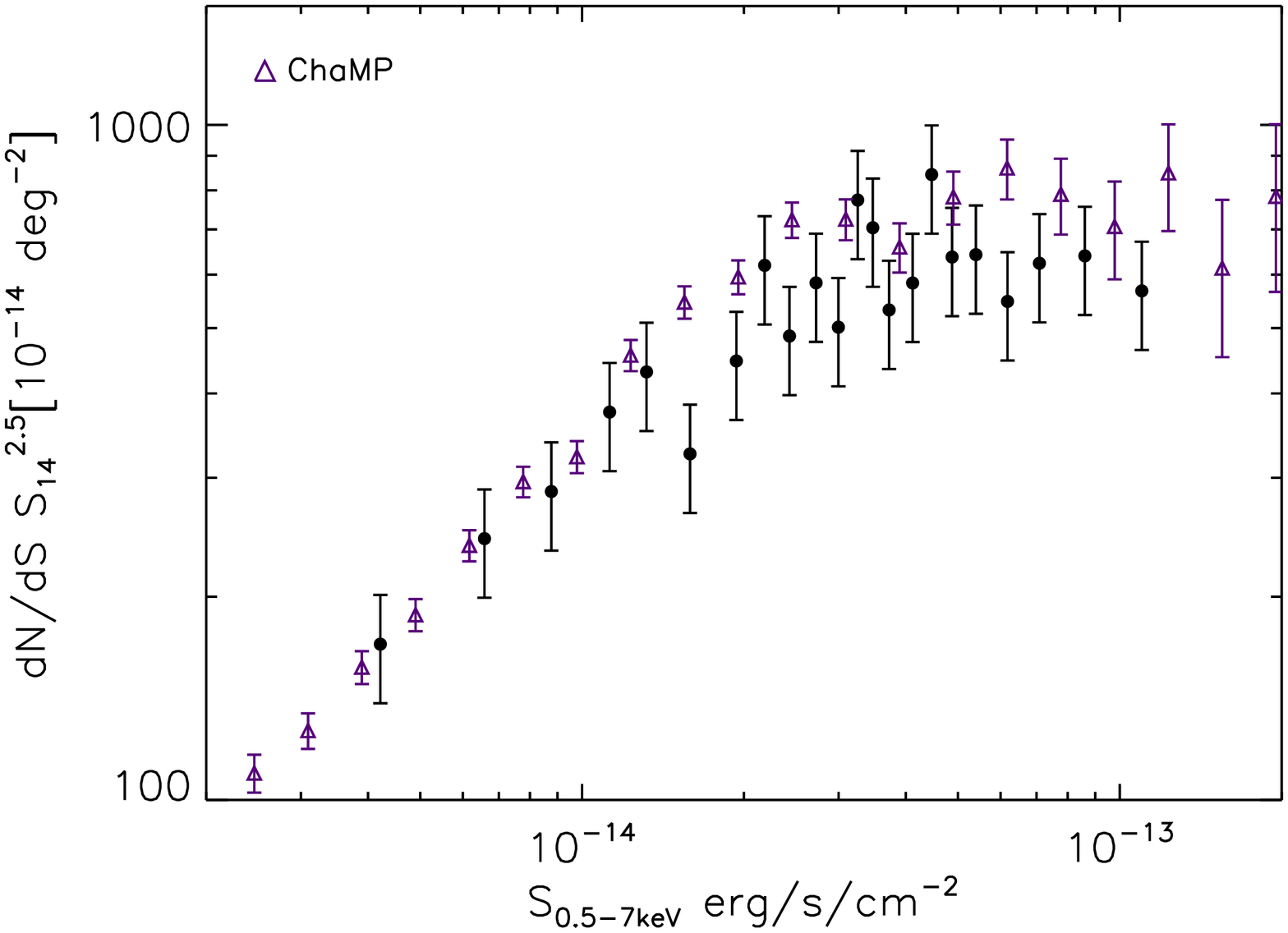}}~
\subfigure[]{\includegraphics[scale=0.35,angle=90]{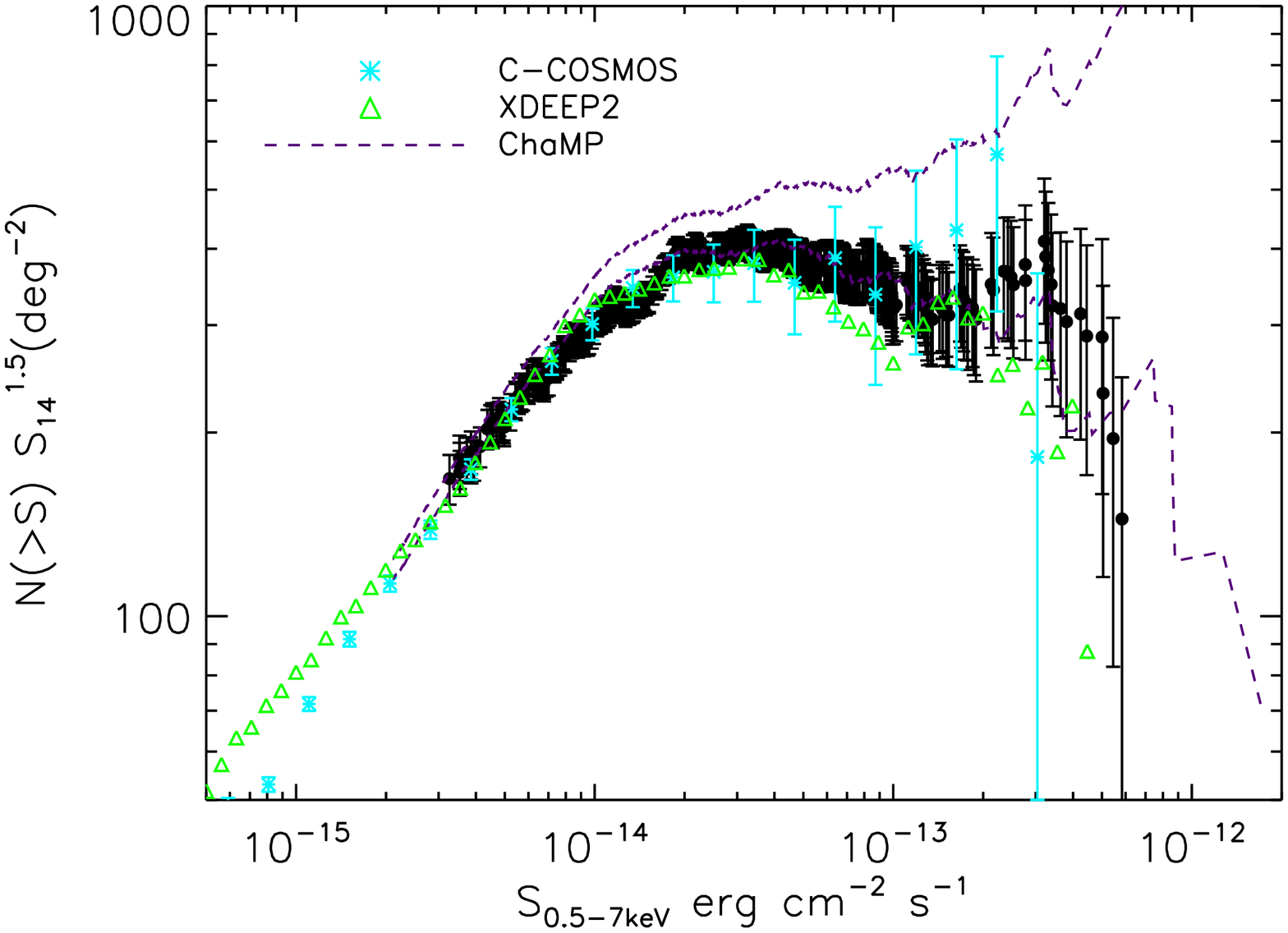}}
\caption[]{\label{logn-logs} Normalized representation of the differential (left) and cumulative (right) number density of X-ray sources as a function of flux for Stripe 82 ACX (filled black circles) compared with other {\it Chandra} surveys, for the soft (top), hard (middle) and full (bottom) X-ray bands. Comparisons include the Extended {\it Chandra} Deep Field South \citep[E-CDFS, red diamonds,][]{leh-ecdfs}, {\it Chandra} COSMOS \citep[C-COSMOS, cyan asterisks,][]{C-Cosmos}, XDEEP2 \citep[green triangles,][]{deep2}, ChaMP \citep[purple triangles, left, 1$\sigma$ confidence interval shown in the purple dashed line, right,][]{champ2} and fit to the XBootes number counts \citep[light blue dash-dot line,][]{kenter}. The dotted line shows the predicted log$N$-log$S$ from the \citet{Gilli} population synthesis model. See text for discussion. }
\end{figure}

 We have imposed a lower sky coverage limit of 0.02 deg$^2$, restricting the Log$N$ - Log$S$ analysis to sources whose fluxes exceed the limiting flux at this sky coverage. The normalized representations of the Stripe 82 ACX Log$N$ - Log$S$ relationships (i.e., dN/dS $\times$ S$_{14}^{2.5}$ and N($>$S) $\times$ S$_{14}^{1.5}$, where S$_{14}$ is flux in units of 10$^{-14}$ erg cm$^{-2}$ s$^{-1}$) are shown in Figure \ref{logn-logs} as black circles, with the binned differential counts on the left and unbinned cumulative integral counts on the right. At the bright flux end, the Stripe 82 ACX number counts are Euclidean, as shown by the horizontal shape in this normalized representation. 

We compare our number counts with previous X-ray surveys that span the range from deep, small area \citep[ECDF-S, at 0.3 deg$^{2}$ with a soft band flux limit of $\sim$1.1$\times$10$^{-16}$ erg cm$^{-2}$ s$^{-1}$,][]{leh-ecdfs}, to slightly larger area, though still rather deep \citep[{\it Chandra}-COSMOS at 0.9 deg$^2$ with a 0.5-2keV flux limit of $\sim$1.9$\times$10$^{-16}$erg cm$^{-2}$ s$^{-1}$,][]{C-Cosmos}, to moderate area and moderate depth \citep[XDEEP2 at 3.2 deg$^2$ with a 0.5-2keV flux limit of $\sim 2\times10^{-16}$erg cm$^{-2}$ s$^{-1}$,][]{deep2} and finally to wide area and moderate depth \citep[ChaMP, at 9.6 deg$^2$ and a soft band flux limit of $\sim$2.5$\times$10$^{-16}$ erg cm$^{-2}$ s$^{-1}$, XBootes at 9 deg$^2$ with a 0.5-7 keV flux limit of $\sim4\times10^{-15}$ erg cm$^{-2}$ s$^{-1}$;][]{champ2, kenter}; only ChaMP (XBootes) has differential number counts (fits) available for comparison. A comparison with the AGN number counts from the population synthesis models from \citet{Gilli} is also presented for the soft and hard bands. We note that although the extended ChaMP/SDSS study has a greater survey area, it represents an optically selected QSO sample incorporating X-ray detections and flux limits, thus no X-ray Log$N$-Log$S$ is published in that work. Stripe 82 ACX, however, considers all X-ray detected sources, regardless of whether or not an optical counterpart has yet been identified.  \citet{georgakakis} also produces Log$N$-Log$S$ for several {\it Chanda} surveys simultaneously, including XBootes and ECDF-S, however as we discuss below, in this analysis it is advantageous for us to consider the surveys independently rather than in aggregate. ECDF-S, C-COSMOS and ChaMP adopt the same spectral model (absorbed power law with $\Gamma$=1.4) to estimate source flux while the XDEEP2 survey uses a spectral index of 1.7. Only XBootes (XDEEP2) defines the hard (and full) energy bands in the same way as Stripe 82 ACX (2-7 keV and 0.5-7 keV, respectively). We have adjusted the values from the remaining comparison surveys using the assumed spectral model of a power law with $\Gamma$=1.4, to shift the hard (2-10 keV for C-COSMOS and the \citet{Gilli} model predictions, 2-8 keV for E-CDFS and ChaMP) and full (0.5-10 keV for C-COSMOS and 0.5-8 keV for ChaMP) bands to be within our defined bandpasses (i.e., we multiplied the comparison fluxes by factors of 0.68, 0.86, 0.76, and 0.91, respectively). 

General agreement exists among the comparison surveys presented here, though ChaMP is systematically higher in all bands and XBootes (XDEEP2) is somewhat lower in the hard (and full) band(s). In the soft and full bands, Stripe 82 ACX agrees with ECDF-S, C-COSMOS, XDEEP2 within the quoted error bars, and is slightly lower than ChaMP (though consistent within the error bars), which has similar (or slightly better) sensitivity at high fluxes. Agreement also exists in the soft band between Stripe 82 ACX and the \citet{Gilli} model. The consistency in the soft band among Stripe 82 ACX and previous {\it Chandra} surveys and the \citet{Gilli} population synthesis model suggests that our method of calculating the soft band flux from the individual ``s'' and ``m'' bands is robust: we do not see a great number of ``missing'' soft sources that would have been identified in a 0.5-2 keV band detection. We note that Stripe 82 ACX is incomplete at low fluxes due to the shallow to moderate depth covered by the majority of archival observations, but the deeper surveys like CDF-S \citep{cdfs} cover this regime better. The characteristic signature of Eddington bias (promotion of spurious sources into the source list due to statistical fluctuations at the detection limit of each observation), which manifests as an ``up-turn'' in the number counts at low fluxes, is absent, likely due to the stringent significance detection threshold we have imposed. This, combined with the consistency between the number counts of Stripe 82 ACX with other surveys, suggests that Eddington bias is negligible in our survey. 

As \citet{deep2} only publish the soft band XDEEP2 Log$N$-Log$S$ and \citet{kenter} do not compare the XBootes differential number counts with other X-ray surveys, the present work is the first to demonstrate a disagreement among {\it Chandra} surveys in the hard band Log$N$-Log$S$. Intriguingly, the Stripe 82 ACX Log$N$-Log$S$ is consistent with XDEEP2 and XBootes in the hard band, yet all three are somewhat lower than other {\it Chandra} surveys. Between 10$^{-14}$ erg cm$^{-2}$ s$^{-1}$ and $8\times10^{-14}$ erg cm$^{-2}$ s$^{-1}$, the 1$\sigma$ error bars between Stripe 82 ACX and ChaMP do not overlap, suggesting that the difference in this range is significant. Though the cumulative Log$N$-Log$S$ data points are not independent, the agreement among Stripe 82 ACX, XBootes and XDEEP2 indicates that the systmatically lower relation we find is not due to several abnormally low data points.

\begin{figure}
\centering
\subfigure[]{\includegraphics[scale=0.35,angle=90]{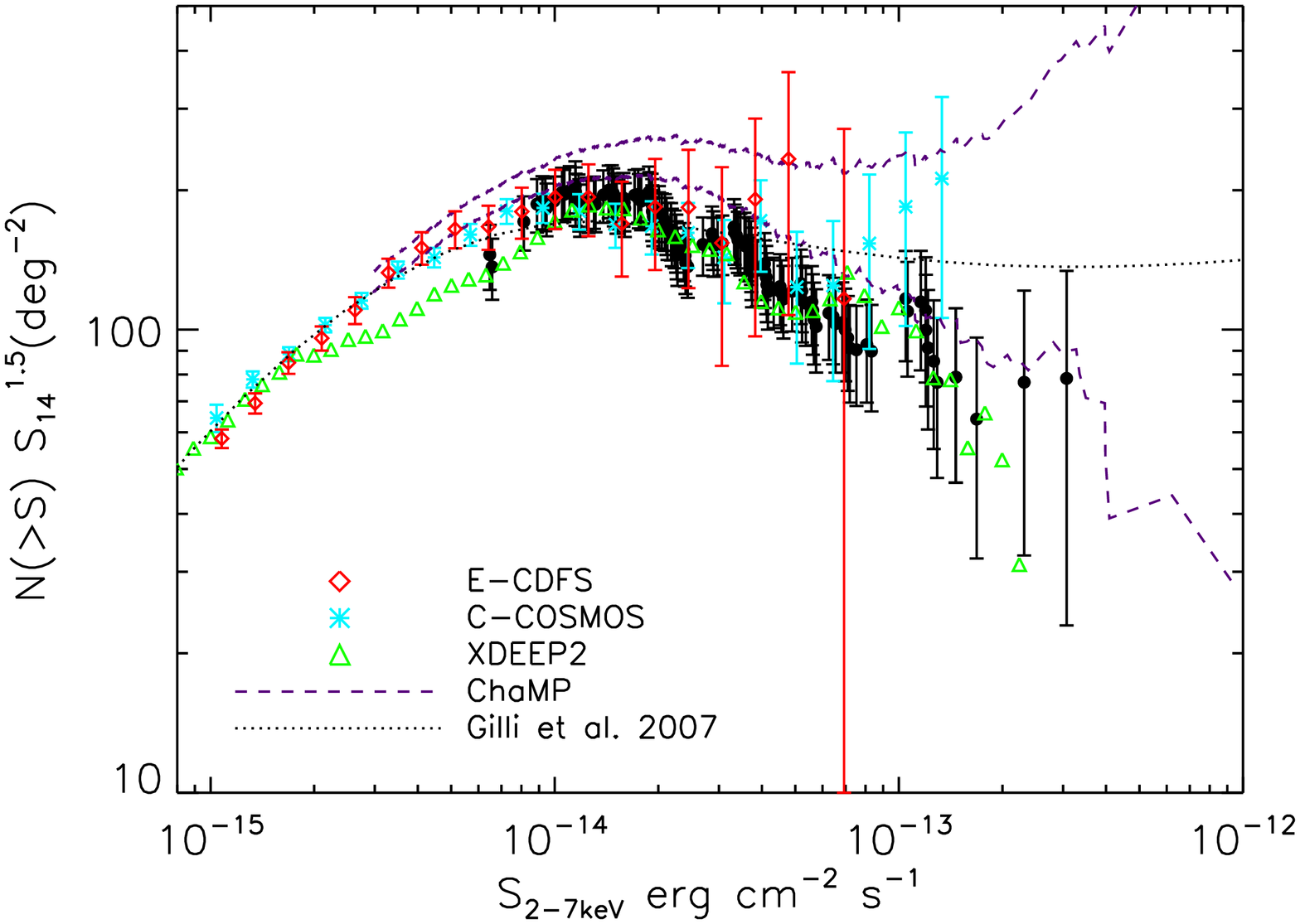}}~
\subfigure[]{\includegraphics[scale=0.35,angle=90]{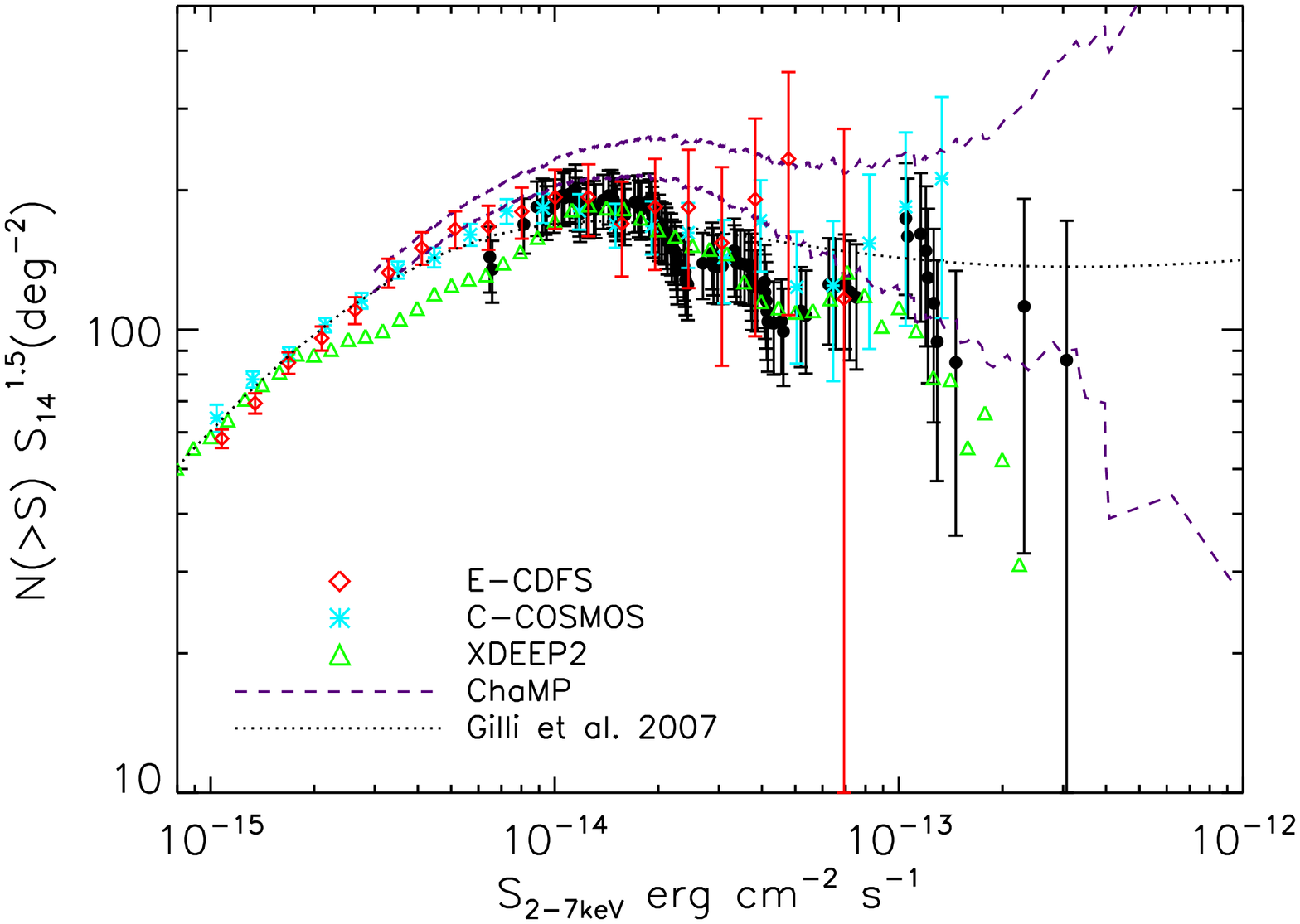}}
\caption[]{\label{hard_test} Hard band number counts in Stripe 82 after removing XDEEP2 pointings (left) and observations with exposure times less than 10 ks (right); survey sensitivity was recalculated to correspond to data used. The Stripe 82 ACX agreement with non-overlapping XDEEP2 files is the same as previously, confirming that our results agree with the XDEEP2 data and analysis. At the same time, removal of short observations ($<$10 ks) increases the normalization slightly, bringing the high flux end into better agreement with ChaMP. However, in both cases a significant disagreement still exists between Stripe 82 ACX and ChaMP within the 2$\times10^{-14}$ - 7$\times10^{-14}$ erg cm$^{-2}$ s$^{-1}$ flux range. We conclude that exposure times can have a significant impact on hard band Log$N$-Log$S$ in {\it Chandra} surveys at high fluxes.}
\end{figure}

What could be the cause of the discrepant normalizations? All three surveys are wide area and are therefore minimally affected by cosmic variance, unlike the smaller area surveys to which we compare our results. However, ChaMP also covers a wide area and is systematically higher, rather than lower, than other {\it Chandra} surveys. To further rule out cosmic variance, and to test whether our results are driven by the inclusion of the XDEEP2 fields, we removed these pointings from the Stripe 82 area, to obtain a completely independent measurement of the hard band Log$N$-Log$S$. The remaining 44 Stripe 82 pointings, covering $\sim$4.3 deg$^2$, are non-contiguous and spread over a wide area, so like ChaMP, these fields provide a more random sampling of the general X-ray population than a contiguous survey. As shown in Figure \ref{hard_test} left, removal of the XDEEP2 fields does not change the normalization of the hard band number counts. Another possibility for the agreement among XBootes, XDEEP2 and Stripe 82 ACX, and their discrepancy with other {\it Chandra} surveys, is the short average exposure time of the former studies: the XBootes pointings are 5 ks in duration, the XDEEP2 survey is a series of 10 ks observations and a majority of the Stripe 82 fields have exposure times under 10 ks. In Figure \ref{hard_test} right, we show the hard band Log$N$-Log$S$ for the $\sim$2 deg$^2$ of observations with exposure time exceeding 10 ks (where the survey sensitivity to account for the subset of pointings has been updated accordingly). The normalization at high fluxes is clearly higher, though a siginficant disagreement still exists betweeen  2$\times10^{-14}$ erg cm$^{-2}$ s$^{-1}$ and 7$\times10^{-14}$ erg cm$^{-2}$ s$^{-1}$. These results indicate that exposure times can play a significant role in determining the source number counts at the high flux end in the hard, but not soft and full bands in {\it Chandra} surveys. However, we note that though Stripe 82 ACX disagrees significantly over a limited flux range with ChaMP in the hard band, it is consistent within the error bars compared to the other 4 surveys we consider.

The full 17 deg$^2$ survey data are needed to generate the high luminosity end of the X-ray QLF, but these Stripe 82 ACX Log$N$-Log$S$ relationships are an important first step towards that goal. An additional area up to 100 deg$^2$ would be ideal for more precise QLF constraints, as well as the ability to study its evolution. For now, as we show below, the current number counts do preferentially trace the high luminosity AGN population and we find many more of these sources in this pilot study than were detected in smaller area X-ray surveys.

\section{Discussion}
Here we investigate the AGN population found in Stripe 82 ACX prior to a time-intensive follow-up spectroscopic campaign. By combining the X-ray catalog with the Sloan Digital Sky Survey (SDSS), we find that over half of the X-ray sources have optical counterparts and half of those have spectroscopic redshifts from existing optical catalogs, enabling identification of these X-ray sources. Below, we comment on this AGN population, comparing the parameter space in which these sources live with those found in smaller area surveys that have extensive spectroscopic follow-up. We also compare the AGN we immediately identify with the population expected from theoretical predictions of AGN population synthesis models.

\subsection{\label{opt_info} Deep Surveys vs. Wide Surveys: Probing a Unique Phase of Black Hole Growth}
Spectroscopic redshifts, or well constrained photometric redshifts, are necessary to robustly calculate X-ray luminosities. The {\it Chandra} Source Catalog provides a matched catalog of X-ray sources with SDSS Data Release 7 \citep{dr7}. Details of the matching algorithm are described in \citet{rots}, who use the Bayesian probabilistic algorithm presented in \citet{budvari}. In brief, for each possible association, a Bayes factor is calculated using a positional uncertainty of 0.1$^{\prime\prime}$ for the SDSS sources and a varying error for {\it Chandra} sources based on the 95\% accuracy limit. A uniform prior is then assumed, which is a function of the number of X-ray and optical sources as well as the number of true pairs. All sources in the cross-matched catalog are those where the probability of a match exceeds 50\%.

Four hundred thirty {\it Chandra} sources have optical counterparts in SDSS, 11 of which have two possible optical counterparts. We have verified these matches by visual inspection and have removed optical counterparts that are saturated, contaminated by proximity to a bright non-point source, coincident with diffraction spikes, or are faint photometric sources from DR7 that do not have photometric detections in Data Release 8 (DR8). For the 11 X-ray objects with two possible optical counterparts, we chose the source with a higher probability of a match (generally the brighter, closer source). This vetting yields 409 matched X-ray and optical sources, with redshifts from the following optical catalogs: SDSS DR8 \citep{dr8}, SDSS-III \citep{sdssiii}, 2SLAQ \citep{2slaq}, WiggleZ \citep{wigglez} and the 4th release of the DEEP2 catalog \citep{deep2_b}. We find spectroscopic redshifts for 204 X-ray sources (186 from SDSS DR8 and DR9, 9 from 2SLAQ, 3 from WiggleZ and 6 from XDEEP2), of which 197 have L$_{x} \geq 10^{42}$ erg s$^{-1}$ (0.5-7 keV), and are therefore likely AGN \citep[e.g.,][]{Persic}.  Here L$_{x}$ represents the observed full band luminosity. Of the 300 X-ray sources not detected in SDSS DR7, 107 were detected by WISE (LaMassa et al., 2013, in prep.). These objects are candidates for elusive obscured high luminosity quasars and will be discussed in a follow-up paper.

\begin{figure}
\centering
{\includegraphics[scale=0.5,angle=90]{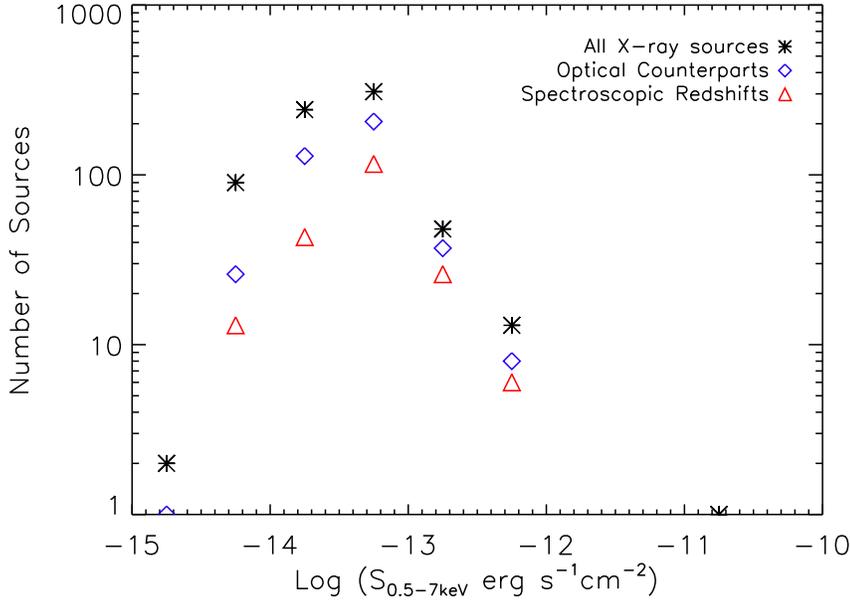}}
\caption[]{\label{frac_opt} Total number of total X-ray sources, X-ray sources with optical (SDSS DR7) counterparts, and X-ray sources with spectroscopic redshift, as a function of full band flux. Optical counterparts and sources with specroscopic redshifts are found at all X-ray flux levels, although the fraction of sources with optical identifications, as well as the fraction with spectroscopic redshifts, increases with increasing X-ray flux. Targets at all X-ray fluxes will be targeted for follow-up optical spectroscopy since the incidence of optical counterparts does not depend strongly on X-ray flux.}
\end{figure}

In Figure \ref{frac_opt}, we plot the total number of X-ray sources and the subset with optical identifications and spectroscopic redshifts as a funcion of full band flux. Interestingly, optical counterparts and spectroscopic identifications are found at all flux levels. The fraction of X-ray sources with optical associations and those with redshifts do increase with full band flux, but a low X-ray flux threshold where optical identifications become improbable is not apparent. Consequently,  follow-up spectroscopy will target sources at all X-ray fluxes.

In Figure \ref{lum_z}, we compare Stripe 82 ACX with the MUSYC survey of CDF-S + E-CDFS \citep{carie} and C-COSMOS \citep{civano2} to illustrate the value of larger areas in expanding the parameter space explored by pencil beam and moderate area surveys. Here we focus only on X-ray-identified AGN with spectroscopic redshifts in all surveys. The spectroscopic completeness between CDF-S + E-CDFS and Stripe 82 ACX  is comparable, $\sim$28\% and $\sim$29\%, respectively (314 of 1134 X-ray sources have reliable spectroscopic redshifts, see Cardamone et al. 2010 for details); C-COSMOS has a higher level of spectroscopic completeness (48\%).  We note that the X-ray luminosities for Stripe 82 ACX and for CDF-S + E-CDFS are observed luminosities, while the C-COSMOS luminosities have been k-corrected to the rest frame.  We also include the predicted AGN luminosity distribution for Stripe 82 ACX using the \citet{treister} population synthesis model where the observed full band 0.5-7 keV area-flux curve was given as the model input. \footnote{An on-line simulator based on the predictions in \citet{treister} is publicly available at http://agn.astroudec.cl/j\_agn/main.html} It is immediately apparent that the AGN identified in Stripe 82 ACX are preferentially at high luminosities (i.e., L$_{x}\geq$ 3$\times 10^{44}$ erg s$^{-1}$) at all redshifts, including in the distant universe at $z > 2$. Stripe 82 ACX uncovers half again as many systems as C-COSMOS beyond log(L$_{x}$) of 44.5 dex (83 vs. 57), although C-COSMOS has a higher level of spectroscopic completeness due to a long-term dedicated follow-up campaign. Though over 100 $z > 3$ AGN have been identified by {\it Chandra}-COSMOS and $z > 3$ quasar luminosity functions have been generated using X-ray data from {\it Chandra}-COSMOS \citep{civano}, {\it XMM-Newton}-COSMOS \citep{brusa10}, and the {\it Chandra} Deep Field South \citep{fiore}, all these sources have faint to moderate luminosities. Wide-area surveys like Stripe 82 ACX are essential to locate rare, L$_{x}> 10^{45}$ erg s$^{-1}$ quasars, which is necessary to populate the high end of the X-ray QLF.

\subsection{Comparison with Predictions}
\begin{figure}
\centering
\subfigure[]{\includegraphics[scale=0.35,angle=90]{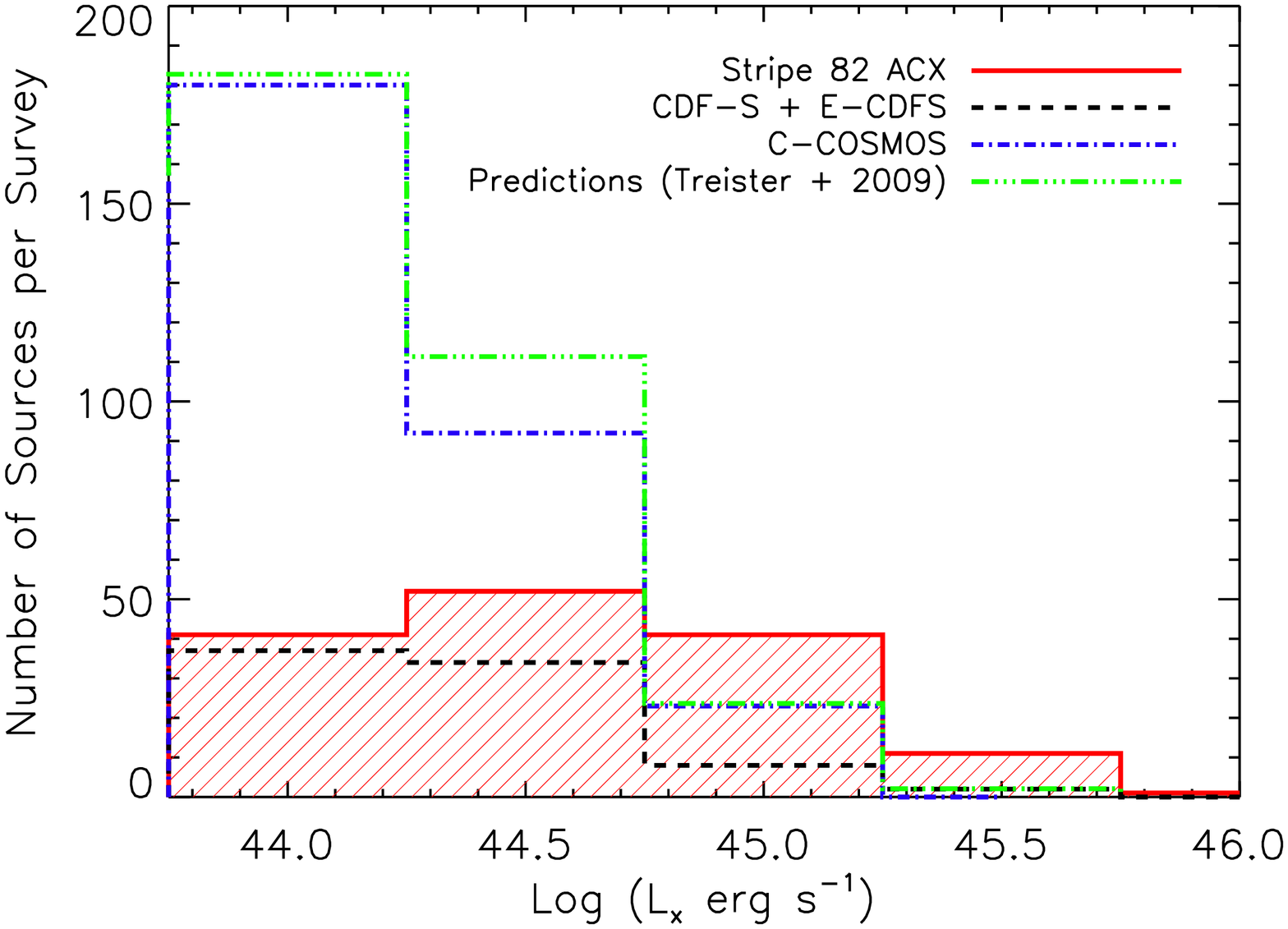}}~
\subfigure[]{\includegraphics[scale=0.35,angle=90]{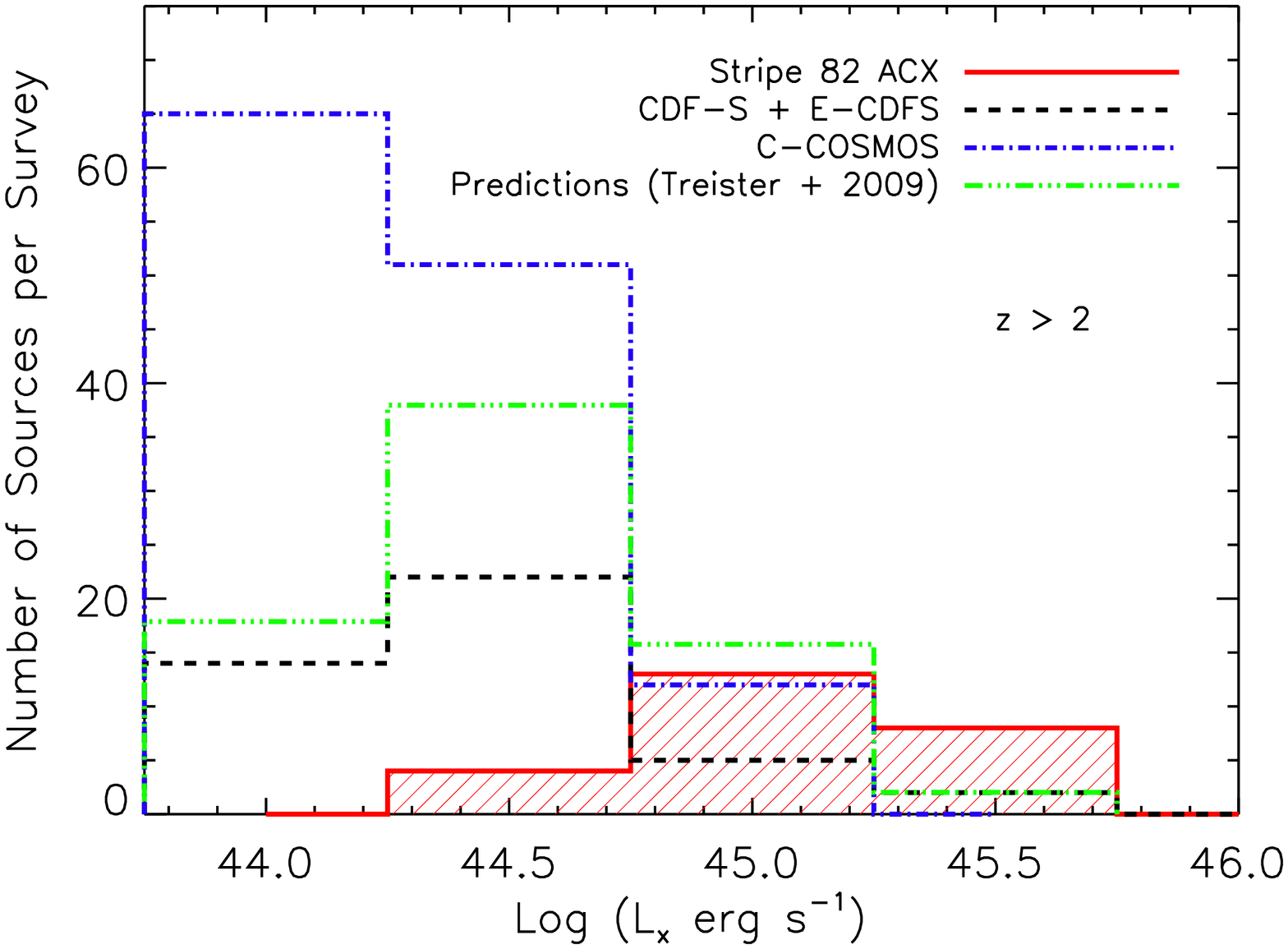}}
\caption[]{\label{lum_z} AGN luminosity distributions in Stripe 82 ACX (red solid line) compared to the smaller area E-CDFS + CDF-S (black dashed line) and C-COSMOS (blue dot-dashed line) surveys for (a) all X-ray sources with spectroscopic redshifts and (b) with $z > 2$. Due to the larger volume probed, Stripe 82 ACX finds more high luminosity AGN, even prior to a dedicated spectroscopic follow-up campaign. Predictions from the \citet{treister} population synthesis models (based on the observed full band area-flux relation) are overplotted (green dot-dot-dash line). At all redshifts, we find more extremely luminous AGN (i.e., L$_X > 3\times10^{45}$ erg s$^{-1}$) than predicted, suggesting that population synthesis models need to be modified at high luminosity and/or high redshift.}

\end{figure}

The \citet{treister} population synthesis model indicates that 580 AGN should be located within the 6.2 deg$^2$ survey area, 137 of which are at high luminosities (L$_{x}\geq$ 3$\times 10^{44}$ erg s$^{-1}$) and 75 beyond a redshift of 2. Figure \ref{lum_z} indicates a majority of the yet-to-be classified AGN are at moderate luminosities. We have found more than half of the expected high luminosity AGN (83 out of 137) and a third of those at $z > 2$ (25 of 75). According to the model predictions, 56 of the $z > 2$ AGN are luminous sources (L$_{x} \geq 44.5$ dex); we have immediately identified half (23) of these.

Our observational results further indicate that population synthesis models can be better refined with our survey data. Only 580 AGN are predicted in Stripe 82 ACX though $\sim$700 X-ray sources lie within this region. Though stars, normal galaxies and X-ray binaries contribute to the total number of X-ray sources, such non-AGN are not numerous enough to explain the discrepancy between the predictions and observations. Additionally, as shown in Figure \ref{lum_z}, more extremely luminous AGN (i.e., L$x > 3\times10^{45}$ erg s$^{-1}$) are found than accounted for in this model. Indeed, the paucity of known AGN at high X-ray luminosities is the reason Stripe 82 X is needed.

\section{Conclusion}
We have presented the first 6.2 deg$^2$ of the current $\sim$17 deg$^2$ X-ray survey covering SDSS Stripe 82. This pilot work concerns the analysis of archival {\it Chandra} observations in the region (Stripe 82 ACX). We utilized source lists and products generated and provided by the {\it Chandra} Source Catalog \citep{CSC}, and the extensive multiwavelength data available in Stripe 82, which greatly streamlined the analysis compared with previous X-ray surveys. Our main results are summarized as follows:

\begin{itemize}

\item Stripe 82 ACX reaches approximate flux limits of 1.2$\times10^{-15}$, 5.4$\times10^{-15}$ and 2.9$\times10^{-15}$ erg s$^{-1}$ cm$^{-2}$, with half area survey coverage at fluxes of 1.5$\times10^{-14}$, 4.8$\times10^{-14}$ and 2.9$\times10^{-14}$ erg s$^{-1}$ cm$^{-2}$, in the soft (0.5-2 keV), hard (2-7 keV) and full (0.5-7 keV) bands, respectively. We detect a total of 709 unique X-ray sources, with 480, 137 and 705 sources at the 4.5$\sigma$ level in the soft, hard and full bands. Of these, 214 were detected solely in the full band and 4 sources detected only in the soft band.

\item The number counts for Stripe 82 ACX are Euclidean at high fluxes. Our soft and full band Log$N$-Log$S$ relations are consistent with the predictions from \citet{Gilli} and with previous {\it Chandra} X-ray surveys: ECDF-S \citep{leh-ecdfs}, C-COSMOS \citep{C-Cosmos}, XDEEP2 \citep{deep2}, XBootes \citep{kenter} and ChaMP \citep{champ2}. We agree with the XDEEP2 and XBootes hard band number counts, but all three are somewhat systematically lower than other surveys. We show that this discrepancy may be due to exposure times, indicating that the length of observations may have a significant impact on the high flux end of the hard band Log$N$-Log$S$ from {\it Chandra} surveys. 

\item 409 {\it Chandra} sources are matched to optical counterparts in the SDSS. We obtained spectroscopic redshifts for 204 of these X-ray objects from SDSS DR8 \citep{dr8}, SDSS-III \citep{sdssiii}, 2SLAQ \citep{2slaq}, WiggleZ \citep{wigglez} and DEEP2 \citep{deep2_b}. One hundred ninety-seven of these objects have X-ray luminosities consistent with AGN (i.e. L$_x \geq 10^{42}$ erg s$^{-1}$); 83 of these are at high luminosity (L$_x \geq 3 \times 10^{42}$ erg s$^{-1}$) and 25 are at $z > 2$. 

\item We compared the AGN luminosity distributions from Stripe 82 ACX with CDF-S + E-CDFS \citep{carie} and C-COSMOS \citep{civano2}; E-CDFS + CDF-S has similar level of spectroscopic completeness with Stripe 82 ACX while C-COSMOS has a somewhat higher level of spectroscopic completeness. This comparison shows that the AGN we immediately identify represent the high luminosity (L$_{x}\geq$ 3$\times 10^{44}$ erg s$^{-1}$) population, at all redshifts, including beyond $z > 2$, while smaller surveys preferentially locate more moderate luminosity AGN.

\item Using population synthesis models from \citet{treister} and the full-band area-flux curve in Stripe 82 ACX, we have demonstrated that more than half of the high luminosity AGN expected to lie in the survey area have been identified prior to any follow-up spectroscopic campaign. We find a third of the high redshift ($z > 2$) AGN and half of the luminous AGN at high redshift that are predicted to lie within Stripe 82 ACX. Slight disagreements between model predictions and our observational results indicate that population synthesis models need to be refined to better accommodate the high luminosity AGN population. This result affirms the importance of a large volume X-ray survey like Stripe 82 X, which will provide important constraints for these models.

\end{itemize}

\section*{Acknowledgments}
We thank the referee for insightful critiques and comments that improved this manuscript. This research has made use of data obtained from the Chandra Source Catalog, provided by the Chandra X-ray Center (CXC) as part of the Chandra Data Archive. We thank Frank Primini and Nina Bonaventura for answering many CSC related questions. We thank Bret Lehmer and Andy Goulding for thoughtful discussions. We thank Bret Lehmer, Andy Goulding and Minsun Kim for access to their LogN-LogS data.


\clearpage

\begin{deluxetable}{llrc}

\tablewidth{0pt}
\tablecaption{\label{sample} {\it Chandra} Observations in the Stripe 82 Field}
\tablehead{

\colhead{ObsID} & \colhead{RA} & \colhead{Dec}    & \colhead{Net Exposure Time} \\
                &              &                  & \colhead{(ks)}  }

\startdata
00344  &   40.670  &  -0.0132  &   47.4  \\
02101  &    7.898  &   0.5724  &    6.7  \\
02115  &  358.223  &  -0.4809  &    5.8  \\
02178  &    9.267  &  -1.1523  &   27.5  \\
02252  &    5.639  &   0.3486  &   71.2  \\
03039  &  334.183  &   0.2301  &    7.4  \\
03576  &   43.217  &  -1.2749  &   39.7  \\
04011  &   40.439  &   0.4431  &    5.0  \\
04100  &   29.209  &   0.8857  &    5.6  \\
04105  &   58.676  &   0.6173  &    9.8  \\
04686  &   40.910  &   1.4036  &    5.7  \\
04825  &   12.526  &  -0.8886  &   13.0  \\
04827  &  359.326  &   0.7306  &   12.7  \\
04829  &    3.275  &   0.0755  &    6.7  \\
04830  &   13.480  &  -0.0526  &    7.1  \\
04832  &   18.115  &  -1.2061  &    5.9  \\
04861  &    0.628  &   0.8331  &    5.7  \\
04963  &   19.723  &  -1.0019  &   39.3  \\
05244  &   38.691  &  -0.8400  &   13.3  \\
05617  &    1.468  &  -0.1155  &   16.9  \\
05650  &   55.807  &   1.0603  &    7.9  \\
05654  &   50.956  &  -0.4971  &    8.3  \\
05694  &   12.541  &  -0.6502  &    7.9  \\
05695  &   49.960  &  -0.9807  &   11.3  \\
05699  &  359.630  &  -0.3740  &    6.3  \\
06128  &  359.225  &  -1.0006  &   17.9  \\
06129  &  359.225  &  -1.0006  &   19.3  \\
06802  &   20.923  &   0.7433  &    9.8  \\
06890  &  331.189  &   0.5283  &    9.4  \\
07241  &   45.300  &  -0.5539  &   49.5  \\
07746  &   14.090  &   0.5433  &    9.9  \\
07747  &   20.134  &  -0.9172  &   10.1  \\
07748  &   23.568  &   0.2371  &    9.9  \\
07749  &   27.385  &  -0.8010  &   10.0  \\
07750  &   29.320  &  -0.8847  &    9.6  \\
07867  &  352.998  &   0.2874  &   21.9  \\
07868  &  352.998  &   0.2874  &   29.8  \\
08173  &  343.778  &   0.9778  &   15.9  \\
08259  &   15.255  &  -0.4123  &   16.8  \\
08601  &  353.253  &   0.2457  &    9.1  \\
08602  &  352.662  &   0.2818  &    8.9  \\
08603  &  353.468  &   0.2078  &    8.8  \\
08604  &  351.640  &   0.2577  &    8.8  \\
08605  &  353.372  &   0.0153  &    8.8  \\
08606  &  352.973  &   0.2194  &    8.8  \\
08607  &  351.899  &   0.2501  &    8.8  \\
08608  &  351.723  &   0.0178  &    8.8  \\
08609  &  353.090  &  -0.0110  &    8.8  \\
08610  &  352.159  &   0.3047  &    8.8  \\
08611  &  352.012  &   0.0234  &    8.8  \\
08612  &  351.474  &   0.0274  &    8.8  \\
08613  &  352.254  &   0.0648  &    8.8  \\
08614  &  352.803  &   0.0608  &    8.8  \\
08615  &  351.481  &   0.2616  &    8.8  \\
08616  &  352.543  &   0.0553  &    8.8  \\
08617  &  352.422  &   0.2914  &    8.8  \\
08619  &   36.640  &   0.7024  &    9.0  \\
08620  &   36.730  &   0.4813  &    8.7  \\
08621  &   36.887  &   0.7025  &    9.1  \\
08622  &   37.144  &   0.7045  &    8.7  \\
08623  &   37.392  &   0.7045  &    8.7  \\
08624  &   37.639  &   0.7027  &    8.7  \\
08625  &   37.890  &   0.7613  &    8.7  \\
08626  &   36.979  &   0.4814  &    8.9  \\
08627  &   37.227  &   0.4739  &    8.5  \\
08628  &   37.483  &   0.4721  &    8.5  \\
08629  &   37.727  &   0.4740  &    8.5  \\
08630  &   37.890  &   0.5974  &    8.5  \\
08960  &  323.372  &  -0.8231  &   11.5  \\
09594  &  343.778  &   0.9778  &   15.0  \\
09719  &  352.998  &   0.2874  &    8.1  \\
10388  &  318.970  &   0.0210  &    9.5  \\
11351  &   36.618  &   1.1605  &    7.5  \\

\enddata

\end{deluxetable}

\clearpage

\begin{landscape}
\begin{deluxetable}{llccll}

\tablewidth{0pt}
\tablecaption{\label{info} Energy Band Summary}
\tablehead{

\colhead{Band} & \colhead{Energy Range} & \colhead{Conversion Factors\tablenotemark{1}}    & \colhead{Number of Sources} & \colhead{Flux Limit} & \colhead{Depth of half survey area}\\
               &                        & \colhead{10$^{-9}$ erg/cm$^{2}$/s/photon}   &                    & \colhead{erg/s/cm$^{2}$} & \colhead{erg/s/cm$^{2}$} }

\startdata

Soft   & 0.5-2 keV & 1.67 & 480 & 1.2$\times10^{-15}$ & 1.5$\times10^{-14}$\\
Hard   & 2-7 keV   & 6.08 & 137 & 5.4$\times10^{-15}$ & 4.8$\times10^{-14}$\\
Full   & 0.5-7 keV & 3.31 & 705 & 2.9$\times10^{-15}$ & 2.9$\times10^{-14}$\\

\enddata

\tablenotetext{1}{Based on absorbed power law model with N$_{H}=3\times10^{20}$ cm$^{-2}$ and $\Gamma$=1.4.}
\end{deluxetable}

\end{landscape}

\clearpage


\clearpage

\clearpage

\clearpage

\clearpage


\begin{thebibliography}{}

\bibitem[\protect\citeauthoryear{Abazajian et al.}{2009}]{dr7} Abazajian, K.~N., Adelman-McCarthy, J.~K., Ag{\"u}eros, M.~A., et al.\ 2009, ApJS, 182, 543 

\bibitem[\protect\citeauthoryear{Aihara et al.}{2011}]{dr8} Aihara, H., Allende Prieto, C., An, D., et al.\ 2011, ApJS, 193, 29 

\bibitem[\protect\citeauthoryear{Alexander et al.}{2003}]{cdfn} Alexander, D.~M., Bauer, F.~E., Brandt, W.~N., et al.\ 2003, AJ, 126, 539 

\bibitem[\protect\citeauthoryear{Becker et al.}{1995}]{Becker} Becker, R.~H., White, R.~L., \& Helfand, D.~J.\ 1995, ApJ, 450, 559 

\bibitem[\protect\citeauthoryear{Bolton et al.}{2012}]{Bolton} Bolton, A.~S., Schlegel, D.~J., Aubourg, E., et al.\ 2012, arXiv:1207.7326 

\bibitem[\protect\citeauthoryear{Bongiorno et al.}{2010}]{bongiorno} Bongiorno, A., Mignoli, M., Zamorani, G., et al.\ 2010, A\&A, 510, A56 

\bibitem[\protect\citeauthoryear{Brusa et al.}{2010}]{brusa10} Brusa, M., Civano, F., Comastri, A., et al.\ 2010, ApJ, 716, 348 

\bibitem[\protect\citeauthoryear{Brusa et al.}{2007}]{brusa} Brusa, M., Zamorani, G., Comastri, A., et al.\ 2007, ApJS, 172, 353 

\bibitem[\protect\citeauthoryear{Budav{\'a}ri \& Szalay}{2008}]{budvari} Budav{\'a}ri, T., \& Szalay, A.~S.\ 2008, ApJ, 679, 301 

\bibitem[\protect\citeauthoryear{Cappelluti et al.}{2009}]{xmm-cap} Cappelluti, N., Brusa, M., Hasinger, G., et al.\ 2009, A\&A, 497, 635 

\bibitem[\protect\citeauthoryear{Cappi et al.}{2006}]{cappi} Cappi, M., Panessa, F., Bassani, L., et al.\ 2006, A\&A, 446, 459 

\bibitem[\protect\citeauthoryear{Cardamone et al}{2010}]{carie} Cardamone, C.~N., van Dokkum, P.~G., Urry, C.~M., et al.\ 2010, ApJS, 189, 270 

\bibitem[\protect\citeauthoryear{Civano et al.}{2012}]{civano2} Civano, F., Elvis, M., Brusa, M., et al.\ 2012, arXiv:1205.5030 

\bibitem[\protect\citeauthoryear{Civano et al.}{2011}]{civano} Civano, F., Brusa, M., Comastri, A., et al.\ 2011, ApJ, 741, 91 

\bibitem[\protect\citeauthoryear{Comastri et al.}{2011}]{comastri} Comastri, A., Ranalli, P., Iwasawa, K., et al.\ 2011, A\&A, 526, L9 

\bibitem[\protect\citeauthoryear{Das et al.}{2011}]{das} Das, S., Marriage, T.~A., Ade, P.~A.~R., et al.\ 2011, ApJ, 729, 62 

\bibitem[\protect\citeauthoryear{Drinkwater et al.}{2010}]{wigglez} Drinkwater, M.~J., Jurek, R.~J., Blake, C., et al.\ 2010, MNRAS, 401, 1429 


\bibitem[\protect\citeauthoryear{Eisenstein et al.}{2011}]{sdssiii} Eisenstein, D.~J., Weinberg, D.~H., Agol, E., et al.\ 2011, AJ, 142, 72 

\bibitem[\protect\citeauthoryear{Elvis et al.}{2009}]{C-Cosmos} Elvis, M., Civano, F., Vignali, C., et al.\ 2009, ApJS, 184, 158 

\bibitem[\protect\citeauthoryear{Elyiv et al}{2012}]{lss2} Elyiv, A., Clerc, N., Plionis, M., et al.\ 2012, A\&A, 537, A131 

\bibitem[\protect\citeauthoryear{Emerson \& Sutherland}{2010}]{vista} Emerson, J., \& Sutherland, W.\ 2010, The Messenger, 139, 2 

\bibitem[\protect\citeauthoryear{Evans et al.}{2010}]{CSC} Evans, I.~N., Primini, F.~A., Glotfelty, K.~J., et al.\ 2010, ApJS, 189, 37 

\bibitem[\protect\citeauthoryear{Fiore et al.}{2012}]{fiore} Fiore, F., Puccetti, S., Grazian, A., et al.\ 2012, A\&A, 537, A16 

\bibitem[\protect\citeauthoryear{Gawiser et al.}{2006}]{musyc} Gawiser, E., van Dokkum, P.~G., Herrera, D., et al.\ 2006, ApJS, 162, 1 

\bibitem[\protect\citeauthoryear{Georgakakis et al.}{2008}]{georgakakis} Georgakakis, A., Nandra, K., Laird, E.~S., Aird, J., \& Trichas, M.\ 2008, MNRAS, 388, 1205 

\bibitem[\protect\citeauthoryear{Giavalisco et al.}{2004}]{giv-goods} Giavalisco, M., Ferguson, H.~C., Koekemoer, A.~M., et al.\ 2004, ApJl, 600, L93 

\bibitem[\protect\citeauthoryear{Gilli et al.}{2007}]{Gilli} Gilli, R., Comastri, A., \& Hasinger, G.\ 2007, A\&A, 463, 79 

\bibitem[\protect\citeauthoryear{Goulding et al.}{2012}]{deep2} Goulding, A.~D., Forman, W.~R., Hickox, R.~C., et al.\ 2012, arXiv:1206.6884 

\bibitem[\protect\citeauthoryear{Green et al.}{2009}]{green} Green, P.~J., Aldcroft, T.~L., Richards, G.~T., et al.\ 2009, ApJ, 690, 644 

\bibitem[\protect\citeauthoryear{Hasinger et al.}{2007}]{xmmcosmos} Hasinger, G., Cappelluti, N., Brunner, H., et al.\ 2007, ApJS, 172, 29 

\bibitem[\protect\citeauthoryear{Hodge et al.}{2011}]{Hodge} Hodge, J.~A., Becker, R.~H., White, R.~L., Richards, G.~T., \& Zeimann, G.~R.\ 2011, AJ, 142, 3 

\bibitem[\protect\citeauthoryear{Kauffmann et al.}{2003}]{kauff} Kauffmann, G., Heckman, T.~M., Tremonti, C., et al.\ 2003, MNRAS, 346, 1055 

\bibitem[\protect\citeauthoryear{Kenter et al.}{2005}]{kenter} Kenter, A., Murray, S.~S., Forman, W.~R., et al.\ 2005, ApJS, 161, 9 

\bibitem[\protect\citeauthoryear{Kewley et al.}{2001}]{kewley} Kewley, L.~J., Dopita, M.~A., Sutherland, R.~S., Heisler, C.~A., \& Trevena, J.\ 2001, ApJ, 556, 121 

\bibitem[\protect\citeauthoryear{Kim et al}{2007a}]{champ1} Kim, M., Kim, D.-W., Wilkes, B.~J., et al.\ 2007a, ApJS, 169, 401 

\bibitem[\protect\citeauthoryear{Kim et al.}{2007b}]{champ2} Kim, M., Wilkes, B.~J., Kim, D.-W., et al.\ 2007b, ApJ, 659, 29 

\bibitem[\protect\citeauthoryear{Kochanek et al.}{2012}]{kochanek} Kochanek, C.~S., Eisenstein, D.~J., Cool, R.~J., et al.\ 2012, ApJS, 200, 8 

\bibitem[\protect\citeauthoryear{Laird et al.}{2009}]{aegisx} Laird, E.~S., Nandra, K., Georgakakis, A., et al.\ 2009, ApJS, 180, 102 

\bibitem[\protect\citeauthoryear{Lawrence et al.}{2007}]{ukidss} Lawrence, A., Warren, S.~J., Almaini, O., et al.\ 2007, MNRAS, 379, 1599 

\bibitem[\protect\citeauthoryear{Lehmer et al.}{2005}]{leh-ecdfs} Lehmer, B.~D., Brandt, W.~N., Alexander, D.~M., et al.\ 2005, ApJS, 161, 21 

\bibitem[\protect\citeauthoryear{Martin et al.}{2005}]{galex} Martin, D.~C., Fanson, J., Schiminovich, D., et al.\ 2005, ApJl, 619, L1 

\bibitem[\protect\citeauthoryear{Mateos et al.}{2008}]{mateos} Mateos, S., Warwick, R.~S., Carrera, F.~J., et al.\ 2008, A\&A, 492, 51 

\bibitem[\protect\citeauthoryear{Murray et al.}{2005}]{murray} Murray, S.~S., Kenter, A., Forman, W.~R., et al.\ 2005, ApJS, 161, 1 

\bibitem[\protect\citeauthoryear{Newman et al.}{2012}]{deep2_b} Newman, J.~A., Cooper, M.~C., Davis, M., et al.\ 2012, arXiv:1203.3192 

\bibitem[\protect\citeauthoryear{Panessa et al.}{2006}]{panessa} Panessa, F., Bassani, L., Cappi, M., et al.\ 2006, A\&A, 455, 173 

\bibitem[\protect\citeauthoryear{Persic et al.}{2004}]{Persic} Persic, M., Rephaeli, Y., Braito, V., et al.\ 2004, A\&A, 419, 849 

\bibitem[\protect\citeauthoryear{Pierre et al.}{2004}]{lss1} Pierre, M., Valtchanov, I., Altieri, B., et al.\ 2004, JCAP, 9, 11 

\bibitem[\protect\citeauthoryear{Primini et al.}{2011}]{Primini} Primini, F.~A., Houck, J.~C., Davis, J.~E., et al.\ 2011, ApJS, 194, 37 

\bibitem[\protect\citeauthoryear{Puccetti et al.}{2009}]{Puccetti} Puccetti, S., Vignali, C., Cappelluti, N., et al.\ 2009, ApJS, 185, 586 

\bibitem[\protect\citeauthoryear{Richards et al.}{2006}]{Richards} Richards, G.~T., Strauss, M.~A., Fan, X., et al.\ 2006, AJ, 131, 2766 

\bibitem[\protect\citeauthoryear{Richards et al.}{2005}]{2slaq} Richards, G.~T., Croom, S.~M., Anderson, S.~F., et al.\ 2005, MNRAS, 360, 839 

\bibitem[\protect\citeauthoryear{Ross et al.}{2012}]{Ross} Ross, N.~P., McGreer, I.~D., White, M., et al.\ 2012, arXiv:1210.6389

\bibitem[\protect\citeauthoryear{Rots \& Budav{\'a}ri}{2011}]{rots} Rots, A.~H., \& Budav{\'a}ri, T.\ 2011, ApJS, 192, 8 

\bibitem[\protect\citeauthoryear{Schawinski et al.}{2007}]{schawinski} Schawinski, K., Thomas, D., Sarzi, M., et al.\ 2007, MNRAS, 382, 1415 

\bibitem[\protect\citeauthoryear{Scoville et al.}{2007}]{cosmos} Scoville, N., Aussel, H., Brusa, M., et al.\ 2007, ApJS, 172, 1 

\bibitem[\protect\citeauthoryear{Treister et al.}{2012}]{treister12} Treister, E., Schawinski, K., Urry, C.~M., \& Simmons, B.~D.\ 2012, ApJL, 758, L39 

\bibitem[\protect\citeauthoryear{Treister et al.}{2009}]{treister} Treister, E., Urry, C.~M., \& Virani, S.\ 2009, ApJ, 696, 110 

\bibitem[\protect\citeauthoryear{Treister et al.}{2004}]{goods} Treister, E., Urry, C.~M., Chatzichristou, E., et al.\ 2004, ApJ, 616, 123 

\bibitem[\protect\citeauthoryear{Trichas et al.}{2012}]{trichas} Trichas, M., Green, P.~J., Silverman, J.~D., et al.\ 2012, ApJS, 200, 17 

\bibitem[\protect\citeauthoryear{Trouille et al}{2011}]{trouille} Trouille, L., Barger, A.~J., \& Tremonti, C.\ 2011, ApJ, 742, 46 

\bibitem[\protect\citeauthoryear{Trouille et al.}{2008}]{clans} Trouille, L., Barger, A.~J., Cowie, L.~L., Yang, Y., \& Mushotzky, R.~F.\ 2008, ApJS, 179, 1 

\bibitem[\protect\citeauthoryear{Virani et al.}{2006}]{vir-ecdfs} Virani, S.~N., Treister, E., Urry, C.~M., \& Gawiser, E.\ 2006, AJ, 131, 2373 

\bibitem[\protect\citeauthoryear{Xue et al.}{2011}]{cdfs} Xue, Y.~Q., Luo, B., Brandt, W.~N., et al.\ 2011, ApJS, 195, 10 

\bibitem[\protect\citeauthoryear{Yang et al}{2004}]{clasxs} Yang, Y., Mushotzky, R.~F., Steffen, A.~T., Barger, A.~J., \& Cowie, L.~L.\ 2004, AJ, 128, 1501 

\end{thebibliography}
\end{document}